\documentclass[twocolumn, groupedaddress, assymb, amsmath]{revtex4}
\usepackage{}
\usepackage{graphicx}
\usepackage{amssymb}
\usepackage{amsmath}

\usepackage{extarrows}
\usepackage[bookmarks=false]{hyperref}
\hypersetup{colorlinks=true, citecolor=blue, linkcolor=blue, urlcolor=blue, pdfstartview=FitH, bookmarksopen=true}

\usepackage{color}
\definecolor{zzz}{rgb}{0.9,0.0,0.4}
\begin{document}
\title{Simultaneous nonreciprocal unconventional photon blockade via two degenerate optical parametric amplifiers in spinning resonators}
\author{J. X. Yang,$^{1}$ Cheng Shang,$^{2, 3, }$\footnote{ \textcolor{zzz}{Corresponding author: cheng.shang@riken.jp}} Yan-Hui Zhou,$^{4, }$\footnote{ \textcolor{zzz}{Corresponding author: yanhuizhou@126.com}} and H. Z. Shen$^{1, }$\footnote{ \textcolor{zzz}{Corresponding author: shenhz458@nenu.edu.cn}}}
\affiliation{$^1$Center for Quantum Sciences and School of Physics,
Northeast Normal University,  Changchun 130024,  China\\
$^2$Analytical quantum complexity RIKEN Hakubi Research Team, RIKEN Center for Quantum Computing (RQC), Wako, Saitama 351-0198, Japan\\
$^3$Department of Physics, The University of Tokyo, 5-1-5 Kashiwanoha, Kashiwa, Chiba 277-8574, Japan\\
$^4$Quantum Information Research Center and Jiangxi Province Key Laboratory of Applied Optical Technology, Shangrao Normal University, Shangrao 334001, China}
\date{\today}

\begin{abstract}
We propose a scheme to achieving simultaneous nonreciprocal unconventional photon blockade in a system of two coupled spining resonators marked by modes $a$ and $b$, each incorporating an degenerate optical parametric amplifier (DOPA). By rotating the resonators, input light from opposite directions induces opposite Sagnac-Fizeau shifts. These shifts result in the emergence or absence of quantum destructive interference in two-photon excitation processes. Specifically, when destructive quantum interference occurs, photons from one input direction are simultaneously blocked in both resonators, whereas the absence of complete destructive quantum interference causes the blockade effect to vanish for inputs from the opposite direction. We analytically give the optimal parameter conditions to achieve simultaneous strong photon blockade with the parametric amplification. By adjusting the Sagnac-Fizeau shifts, we can make mode $a$ nonreciprocal photon blockade, while mode $b$ exhibits photon blockade in both directions. This work lays a theoretical foundation for the development of multimode simultaneous nonreciprocal unconventional single-photon devices, which hold promising potential in multichannel topological optics and chiral quantum technologies.
\end{abstract}

%\pacs{42.50.Pq, 42.50.Ct, 42.50.Ar, 42.50.Dv} \maketitle
%42.50.pq: Cavity quantum electrodynamics; micromasers
%42.50.Ct: Quantum description of interaction of light and matter; related experiments
%42.50.Ar: Photon statistics and coherence theory
%42.50.Dv: Quantum state engineering and measurements

%Ò×

\maketitle
\section{Introduction}

Photon blockade (PB), a process where the presence of one photon inhibits the injection of a second, serves as a fundamental technique for generating single-photon sources \cite{Shields2052007, Ghosh0136022019, Huang21004302022, Tang47052025, Zuo0437152022, Liu0637052024, Zhang0637052024, Zou0537102020, Xu0638532016, Guo0137052022, Li0437072024, Ghosh0136022019, Li24003742025, Lu0136022025, Sun0437152023}. This phenomenon, characterized by photon antibunching and sub-Poissonian photon-number statistics, plays a crucial role in quantum technologies, enabling applications such as interferometers \cite{Gerace2812009}, single-photon transistors \cite{Chang8072007}, and nonclassical isolators \cite{Sayrin0410362015, Tang0438332019}. PB is generally achieved through two mechanisms: conventional photon blockade (CPB) and unconventional photon blockade (UPB). CPB arises from the anharmonic energy levels of a system, preventing transitions to higher-photon states when the single-photon state is occupied \cite{Zhao0638382020}. While CPB requires strong nonlinearities \cite{Feng123042023, Bamba0218022011} or robust coupling with nonlinear elements like atoms, it has been experimentally realized in systems such as cavity quantum electrodynamics (QED) \cite{Birnbaum872005, Reinhard932012, Muller2336012015, Peyronel572012,Zhu0638422017,Lin0538502019,Hou0638172019,Ebrahimi562023} and circuit QED \cite{Lang2436012011, Hoffman0536022011, Faraon8592008}. In contrast, UPB relies on quantum destructive interference between different excitation pathways to suppress multi-photon states, significantly relaxing the need for strong system parameters \cite{Bamba0218022011,Xu0438222014,Gerace0318022014,Tang92522015,Flayac0538102017,Wang2404022021,Shen0437142024,Zuo220202024}. This approach has been successfully demonstrated in platforms like coupled quantum dot-cavity systems \cite{Snijders0436012018}, optical cavities, and superconducting resonators \cite{Vaneph0436022018}. Over the past decades, substantial experimental and theoretical advancements have been made in PB, first observed in an optical cavity coupled to a single two-level atom \cite{Hamsen1336042017}. The phenomenon has since been explored in diverse systems, including cavity optomechanics \cite{Wang20642020,Shi82018,Yuan222022,Zheng0138042019}, magnomechanical devices \cite{Zhang117732021}, and spinning resonators \cite{Huang1536012018}. Importantly, UPB has been realized in weakly nonlinear systems, such as coupled cavities with second- or third-order nonlinearities \cite{Zhou0238382015,Shen0638082015,Ferretti0250122012,Flayac0338362013,Kyriienko0638052014,Liu23004222024,Zhang23001872023,Shen328352015}, Gaussian squeezed
states \cite{Lemonde0638242014,Shen0238562018,Sarma0138262018} and gain cavities \cite{Zhou0438192018}. The realization of two-photon or photon-pair blockade has also been verified in different systems \cite{Miranowicz0238092013,Zhai276492019,Wang26042020,Li0437022024,Qiao0537022024,Feng0435092021,Bin0438582018,Kudlaszyk0538572019}. These studies continue to deepen our understanding of PB and its potential for advancing quantum technologies.

Optical nonreciprocity, which permits light to propagate in one direction while blocking it in the opposite direction, has broad applications in optical isolators \cite{Sayrin0410362015,Tang0438332019}, circulators \cite{Kamal3112011}, unidirectional amplifiers \cite{Shen17972018,Malz0236012018}, noise-free information processing, and invisible sensing \cite{Sounas7742017}. Leveraging this principle, researchers have explored a range of phenomena, including nonreciprocal entanglement \cite{Jiao0640082022,Jiao1436052020,Ren11252022,Chen0241052023}, phonon and magnon lasers \cite{Xu0535012021,Huang33112022,Huang1042012023,Xu52762021,He435062023,Wang49872024}, nanoparticle sensing \cite{Jing14242018}, slow light \cite{Mirza255152019,Peng0335072023}, optical solitons \cite{Li0535222021}, and photon blockade. Nonreciprocal photon blockade (NPB) \cite{Huang1536012018,Jing0337072021,Zhang2403132023,Yuan0535262024}, where photons are blockaded in one direction but not the other, has gained particular attention. The first demonstrations of NPB relied on spinning resonators exploiting the optical Sagnac effect to achieve nonreciprocal conventional photon blockade (NCPB) \cite{Huang1536012018,Xue44242020} through strong nonlinearity or nonreciprocal unconventional photon blockade (NUPB) \cite{Li6302019,Hou51452025,Shen0138262020,Xia0637132021,Xia79072022,Wang640032021} via quantum destructive interference with weak nonlinearity. These effects have been observed in diverse systems, including spinning Kerr resonators, optomechanical setups \cite{Xu1432020,Shang1152022021,Liu128472023}, coupled atom-cavity systems \cite{Xue44242020,Jing0337072021,Gu0437222022,Zhang2403132023,Liu0637012023}, nonlinear cavities, and cavities with parametric amplification. Recent studies have proposed simultaneous NPB \cite{Liu0637012023,Gou0437232023,Zhang0237232024} in two resonators influenced by Fizeau drag effects, while others have suggested utilizing directional parametric amplification for simplifying experimental realizations. These advancements underscore the growing potential of NPB in enabling cutting-edge quantum and photonic technologies.

This paper explores the phenomenon of simultaneous nonreciprocal unconventional photon blockade in a system of two coupled spinning resonators each containing a degenerate optical parametric amplifier (DOPA). The rotation of the resonators introduces opposite Sagnac-Fizeau shifts for inputs from opposite ends, resulting in the emergence or absence of destructive quantum interference in multiple excitation pathways inducing two-photon states within the cavities. For one input direction, destructive interference causes photon blockade to occur simultaneously in both resonators, while for the other direction, the blockade vanishes due to incomplete interference. We analytically derive optimal conditions for achieving strong photon antibunching in both resonators, verifying the results through numerical simulations. Our findings reveal that simultaneous nonreciprocal unconventional photon blockade depends critically on the coupling strength and nonlinear gain, and can be finely controlled by adjusting the probe field amplitude and the driving field applied to the DOPA. This scheme offers a pathway for realizing simultaneous nonreciprocal single-photon sources in two modes, with potential applications in multichannel topological systems and chiral quantum technologies.

The present paper is organized as follows. In Sec.~\ref{Sec2}, we present the Hamiltonian of the system. In Sec.~\ref{Sec3}, we provide a detailed derivation of the analytical and numerical solutions for the photon blockade effect. In Sec.~\ref{Sec4}, we investigates the simultaneous NPB of two resonators. Finally, we give a conclusion in Sec.~\ref{Sec5}.

\section{MODEL}\label{Sec2}
\begin{figure}[t]
\centerline{
\includegraphics[width=8.5cm, height=4.5cm, clip]{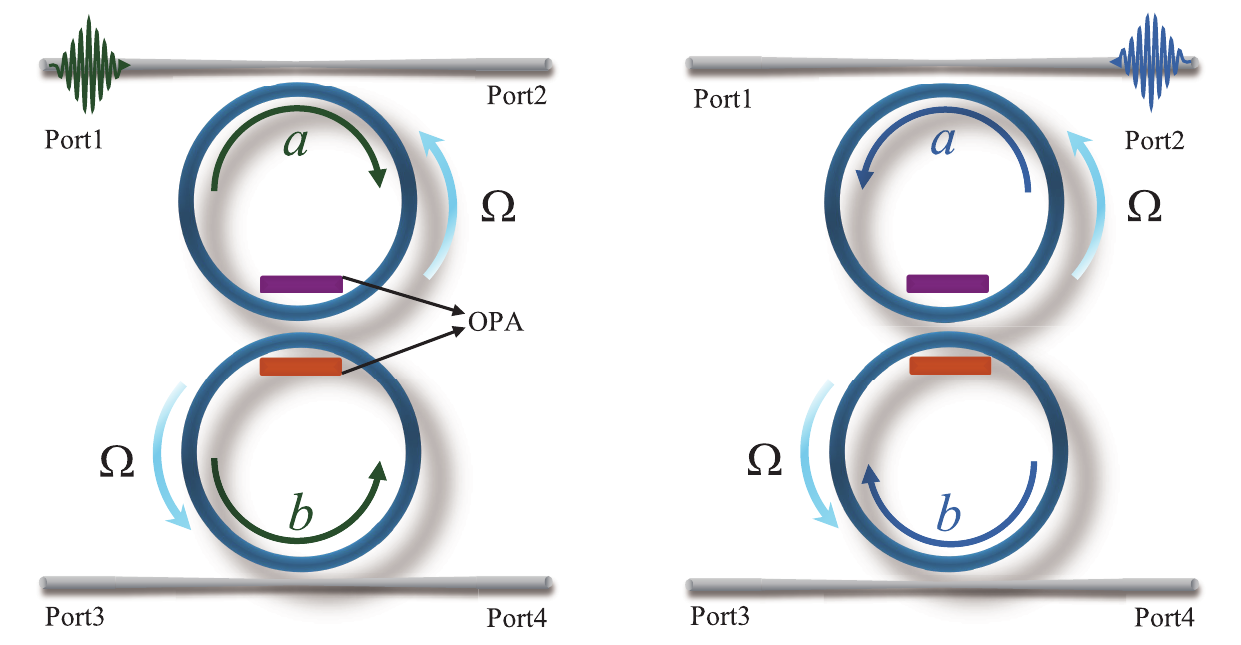}}
\vspace{-0.1cm}
\caption{(Color online) Nonreciprocal photon blockade in two coupled spinning resonators. The purple and orange squares represent degenerate optical parametric amplifiers in the two cavities.  Both cavities rotate counterclockwise at an angular velocity $\Omega$. Here, forward pumping is indicated in green, and backward pumping is shown in blue.}\label{model}
\end{figure}

As shown in Fig.~\ref{model}, we consider two spinning optical resonators, rotating in a counterclockwise rotation with fixed angular velocity $\Omega $ are coupled to each other
at a coupling strength $J$. For light circulating in a spinning resonator, its optical mode experiences a Sagnac-Fizeau shift \cite{Malykin12292000}, i.e., ${\omega _c} \to {\omega _c} \pm {\Delta _{SF}}$ with
\begin{equation}\begin{aligned}
{\Delta _{SF}} =  \Omega \frac{{nR{\omega _c}}}{c}( {1 - \frac{1}{{{n^2}}} - \frac{\lambda }{n}\frac{{dn}}{{d\lambda }}} ),\label{DeltaSF}
\end{aligned}\end{equation}
where $n$ and $R$ are the refractive index of the material and radius of resonator, $c(\lambda )$ is the speed (wavelength) of light in vacuum. The dispersion term $dn/d\lambda $ originates from the relativistic correction of the Sagnac effect, is relatively small, and can be ignored in typical materials. Note that a positive Sagnac-Fizeau shift $({\Delta _{SF}} > 0)$ corresponds to the forward pump case, while a negative shift $({\Delta _{SF}} < 0)$ denotes the backward pump.
Two DOPAs are connected to resonators $a$ and $b$, providing two-photon driving on modes $a$ and $b$. In a rotating frame $\hat V = \exp [ - i{\omega _l}t({{\hat a}^\dag }{\hat a} + {{\hat b}^\dag }{\hat b})]$, the total Hamiltonian of the system is written as $( {\hbar  \equiv 1} )$
\begin{equation}
\begin{aligned}
\hat H =& ({\Delta _a} + {\Delta _{SF}}){{\hat a}^\dag }{\hat a} + ({\Delta _b} - {\Delta _{SF}}){{\hat b}^\dag }{\hat b} + J({{\hat a}^\dag }{\hat b} + {{\hat b}^\dag }{\hat a}) \\
    &+\zeta ({e^{i\varphi }{{\hat a}^\dag }} +{e^{-i\varphi } {\hat a}}) + K({e^{ - i\theta }{{\hat a}^{\dag 2}}} + {e^{i\theta }{{\hat a}^2}}) \\
    &+ F({{\hat b}^{\dag 2}} + {{\hat b}^2}),\label{H}
\end{aligned}
\end{equation}
where $\hat a$ and $\hat b$ are the annihilation operators of the clockwise and counterclockwise cavity modes, respectively. The detunings are defined as ${\Delta _j} = {\omega _j} - {\omega _l}$, where $j=a, b$. $\zeta$ and $\varphi$ are the amplitude and phase of the weak probe field (frequency $\omega_l$) on mode $a$, the fifth term represents
the coupling between the cavity field and the DOPA with the parametric gain $K$, depending on the power of the pump driving the DOPA, and the phase $\theta$ of the pump driving the DOPA. $F$ is the nonlinear gain, which is proportional to the amplitude of the laser field (frequency $\omega_p$) applied on the DOPA.
%figure 1
%and backward pumping is shown in blue.%
\section{PHOTON BLOCKADE}\label{Sec3}
%The statistic properties of the photon will be described by the equal-time second-order correlation function in the steady state, which can be obtained through analytically solving the non-Hermitian Schr\"{o}dinger equation and numerically simulating the quantum master equation.
\subsection{Analytical solution}
The non-Hermitian Hamiltonian is given by adding phenomenologically the decay rate, and its form is as follows
\begin{equation}
\begin{aligned}
{\hat H_{{\rm{nm}}}} = \hat H - \frac{{i{\kappa _a}}}{2}{{\hat a}^\dag }{\hat a} - \frac{{i{\kappa _b}}}{2}{{\hat b}^\dag }{\hat b}.\label{Hnm}
\end{aligned}
\end{equation}
In the weak-drive regime ($\zeta  \ll {\kappa _a} = {\kappa _b} = \kappa $), the Hilbert space of system can be truncated to the low-excitation subspace (up to two photons), and the wave function can be written as
\begin{equation}
\begin{aligned}
\left| \psi  \right\rangle  =& {C_{00}}\left| {00} \right\rangle  + {C_{10}}\left| {10} \right\rangle  + {C_{01}}\left| {01} \right\rangle \\
 &+ {C_{20}}\left| {20} \right\rangle  + {C_{11}}\left| {11} \right\rangle  + {C_{02}}\left| {02} \right\rangle ,\label{Phi}
\end{aligned}
\end{equation}
where ${C_{ab}}$ for $m,n=0,1,2$ are the probability amplitudes corresponding to the bare state $\left| {mn} \right\rangle $. By substituting the quantum state $\left| \psi  \right\rangle $ and non-Hermitian Hamiltonian Eq.~(\ref{Hnm}) into the Schr\"{o}dinger equation $i\partial \left| \psi  \right\rangle /\partial t = {\hat H_{{\rm{nm}}}}\left| \psi  \right\rangle $, we obtain a set of dynamic equations for the probability amplitudes
\begin{equation}
\begin{aligned}
i{{\dot C}_{00}} &= \zeta {e^{ - i\varphi }}{C_{10}} + \sqrt 2 K{e^{i\theta }}{C_{20}} + \sqrt 2 F{C_{02}},\\
i{{\dot C}_{10}} &= (\bar \Delta   + {\Delta _{SF}}){C_{10}} + J{C_{01}} + \zeta {e^{i\varphi }}{C_{00}} + \sqrt 2 \zeta {e^{-i\varphi }}{C_{20}},\\
i{{\dot C}_{01}} &= (\bar \Delta   - {\Delta _{SF}}){C_{01}} + J{C_{10}} + \zeta {e^{ - i\varphi }}{C_{11}},\\
i{{\dot C}_{20}} &= 2(\bar \Delta   + {\Delta _{SF}}){C_{20}} + \sqrt 2 J{C_{11}} + \sqrt 2 \zeta {e^{i\varphi }}{C_{10}} \\
&\quad + \sqrt 2 K{e^{ - i\theta }}{C_{00}},\\
i{{\dot C}_{11}} &= 2\bar \Delta  {C_{11}} + \sqrt 2 J{C_{20}} + \sqrt 2 J{C_{02}} + \zeta {e^{i\varphi }}{C_{01}},\\
i{{\dot C}_{02}} &= 2(\bar \Delta   - {\Delta _{SF}}){C_{02}} + \sqrt 2 J{C_{11}} + \sqrt 2 F{C_{00}},\label{emofpa}
\end{aligned}
\end{equation}
where we assume ${\Delta _1} = {\Delta _2} = \Delta $, $\bar \Delta   = \Delta  - i\kappa /2$. With the condition ${C_{00}} \simeq 1 \gg {C_{10}},{C_{01}} \gg {C_{20}},{C_{11}},{C_{02}}$, By neglecting higher-order terms in each equation enables the analytical derivation of steady-state solutions for the system.
\begin{figure}[t]
\centerline{
\includegraphics[width=8cm, height=3.6cm, clip]{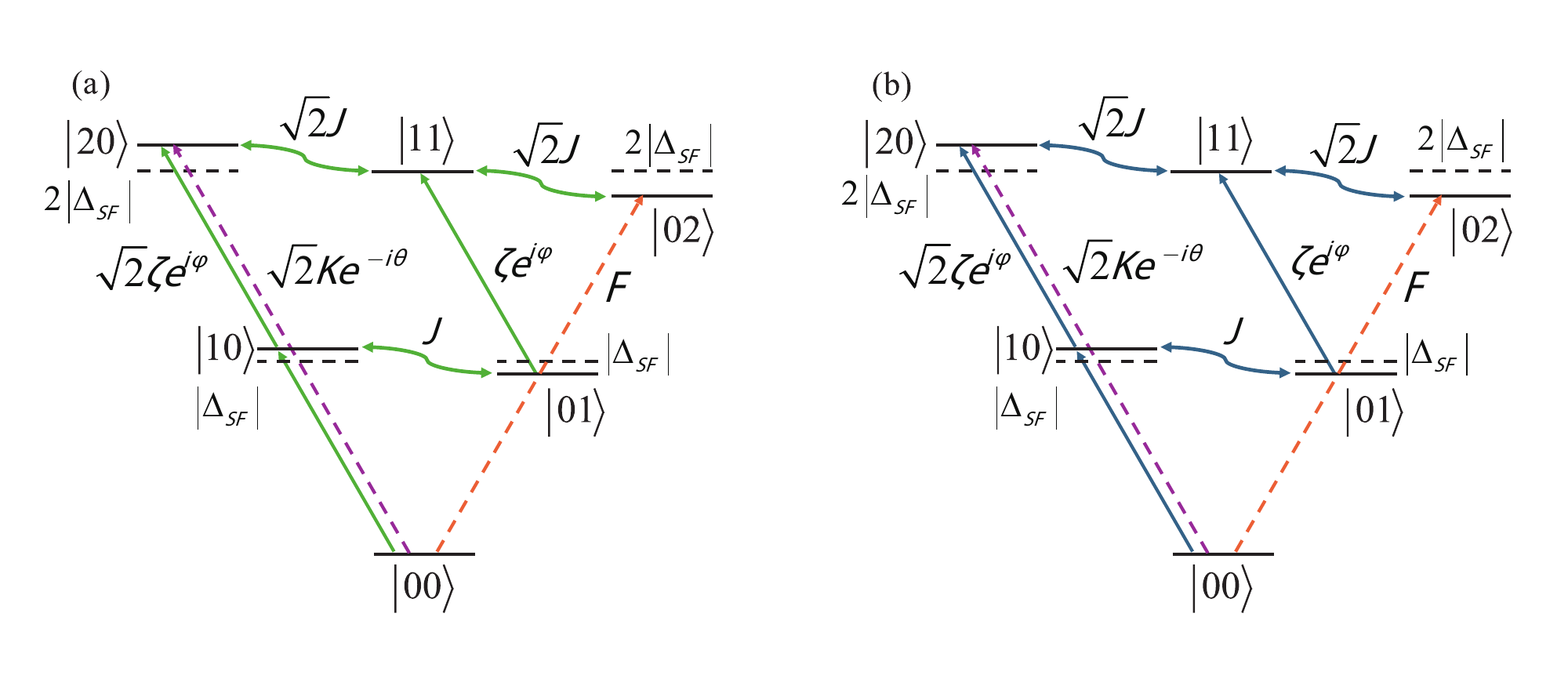}}
\vspace{-0.1cm}
\caption{(Color online) Energy-level diagram of the system. In the cases of two different inputs, the black solid line represents the zero, one, and two-photon states, while the orange and purple solid lines respectively represent the DOPA in cavities $a$ and $b$. The green and blue arrowed lines represent the transmission paths for two input directions. Due to the different transmission environments, destructive quantum interference occurs, preventing the appearance of the two-photon state.}\label{Energylevel}
\end{figure}
\begin{widetext}
\begin{equation}
\begin{aligned}
{C_{10}} =& \frac{{2{e^{i\varphi }}\zeta B}}{A},\\
{C_{01}} =& \frac{{ - 4{e^{i\varphi }}\zeta J}}{A},\\
{C_{20}} =& \frac{{\sqrt 2 {e^{ - i\theta }}(2{e^{i(\theta  + 2\varphi )}}{\zeta ^2}(2\Delta  - i\kappa ){{B}^2} + A(2{e^{i\theta }}F{J^2} - G(2{J^2} - 4{\Delta ^2} + 4\Delta {\Delta _{SF}} + 4i\kappa  - 2i{\Delta _{SF}}\kappa  + {\kappa ^2}))}}{{(2\Delta  - i\kappa ){A^2}}},\\
{C_{11}} =& \frac{{2J( - 4{e^{2i\varphi }}{\zeta ^2}(2\Delta  - i\kappa )B - {e^{ - i\theta }}AK(2\Delta  - 2{\Delta _{SF}} - i\kappa ) + {e^{i\theta }}F(2(\Delta  + {\Delta _{SF}}) - i\kappa ))}}{{(2\Delta  - i\kappa ){A^2}}},\\
{C_{02}} =& \frac{{\sqrt 2 {e^{ - i\theta }}(8{e^{i(\theta  + 2\varphi )}}{J^2}{\zeta ^2}(2\Delta  - i\kappa ) - A( - 2{J^2}K + {e^{i\theta }}F(2{J^2} - 4\Delta (\Delta  + {\Delta _{SF}}) + 2i(2\Delta  + {\Delta _{SF}})\kappa  + {\kappa ^2})))}}{{(2\Delta  - i\kappa ){A^2}}},\label{sssofpa}
\end{aligned}
\end{equation}
\end{widetext}
where $A=4({J^2} - {\Delta ^2} + \Delta _{SF}^2) + 4i\Delta \kappa  + {\kappa ^2}$, $B=2\Delta  - 2{\Delta _{SF}} - i\kappa$.
%The probabilities of finding $m$ particles in the CW mode and $n$ particles in the CCW mode are given by
%\begin{equation}
%\begin{aligned}
%{P_{mn}} = |{C_{mn}}{|^2}.\label{P}
%\end{aligned}
%\end{equation}
%Based on this, we can finally derive the results for the second-order correlation functions:
Then the equal-time second-order photon correlation function can be expressed as
\begin{eqnarray}
\begin{aligned}
g_a^{(2)}(0) \simeq \frac{{2|{C_{20}}{|^2}}}{{|{C_{10}}{|^4}}},\\
g_b^{(2)}(0) \simeq \frac{{2|{C_{02}}{|^2}}}{{|{C_{01}}{|^4}}}.\label{jxg20}
\end{aligned}
\end{eqnarray}

\subsection{Numerical simulation}
To validate the consistency of our analytical solutions, we perform numerical simulations of the system's quantum dynamics. Accounting for microwave cavity dissipation, the open quantum dynamics are governed by the quantum master equation \cite{Scully1997,Walls1994}
\begin{equation}
\begin{aligned}
\dot{\hat{\rho}}  =  - i[\hat H,\hat \rho ] + \frac{\kappa }{2}{\cal D}(\hat a)\hat \rho  + \frac{\kappa }{2}{\cal D}(\hat b)\hat \rho,\label{rho}
\end{aligned}
\end{equation}
where the Lindblad superoperator is denoted as ${\cal D}(\hat x)\hat \rho  = 2\hat x\hat \rho {\hat x^\dag } - {\hat x^\dag }\hat x\hat \rho  - \hat \rho {\hat x^\dag }\hat x$ with the system operator $\hat x$. Here, ${\hat \rho }$ is the system density operator. To analyze second-order correlation functions in the steady-state regime, we first compute the steady-state density operator
${{\hat \rho }_s}$, obtained by solving the stationary condition $\partial \hat \rho /\partial t = 0$. The equal-time second-order correlation function \cite{Glauber25291963} is obtained by
\begin{equation}
\begin{aligned}
g_o^{(2)}(0) = \frac{{\left\langle {{\hat o^\dag }{\hat o^\dag }\hat o\hat o} \right\rangle }}{{{{\left\langle {{\hat o^\dag }\hat o} \right\rangle }^2}}} = \frac{{{\rm{Tr}}({\hat o^\dag }{\hat o^\dag }\hat o\hat o{\hat \rho _s})}}{{{{{\rm{[Tr(}}{\hat o^\dag }\hat o{\hat \rho _s})]}^2}}}.\label{szg20}
\end{aligned}
\end{equation}
The second-order photon correlation function $g_o^{(2)}(0)$ serves as a fundamental metric in both experimental and theoretical investigations of photon blockade phenomena. Here, $g_o^{(2)}(0) < 1$ correspond to photon antibunching with sub-Poissonian statistics, while $g_o^{(2)}(0) > 1$ describes the photon bunching with super-Poissonian statistics.

\section{SIMULTANEOUS NONRECIPROCAL UNCONVENTIONAL PB}\label{Sec4}
By observing Fig.~\ref{Energylevel}, it can be found that there are four paths to access the two-photon state $\left| {20} \right\rangle $: $|00\rangle \xrightarrow{\zeta e^{i\varphi}} |10\rangle \xrightarrow{\sqrt{2}\zeta e^{i\varphi}} |20\rangle$, $|00\rangle \xrightarrow{\sqrt{2} K e^{-i\theta}} |20\rangle$, $|00\rangle \xrightarrow{\zeta e^{i\varphi}} |10\rangle \xrightarrow{J} |01\rangle \xrightarrow{\zeta e^{i\varphi}} |11\rangle \xrightarrow{\sqrt{2}J} |20\rangle$ and $|00\rangle \xrightarrow{F} |02\rangle \xrightarrow{\sqrt{2}J} |11\rangle \xrightarrow{\sqrt{2}J} |20\rangle$. Similarly, there are four paths leading to the state $\left| {02} \right\rangle$: $|00\rangle \xrightarrow{F} |02\rangle$, $|00\rangle \xrightarrow{\zeta e^{i\varphi}} |10\rangle \xrightarrow{\sqrt{2}\zeta e^{i\varphi}} |20\rangle \xrightarrow{\sqrt{2}J} |11\rangle \xrightarrow{\sqrt{2}J} |02\rangle$, $|00\rangle \xrightarrow{\sqrt{2} K e^{-i\theta}} |20\rangle \xrightarrow{\sqrt{2}J} |11\rangle \xrightarrow{\sqrt{2}J} |02\rangle$ and $|00\rangle \xrightarrow{\zeta e^{i\varphi}} |10\rangle \xrightarrow{J} |01\rangle \xrightarrow{\zeta e^{i\varphi}} |11\rangle \xrightarrow{\sqrt{2}J} |02\rangle$. Since there are multiple paths to access the two-photon $|20\rangle$ and $|02\rangle$, it implies that they may simultaneously experience destructive interference under optimized parameters, i.e., ${C_{20}}=0$ and ${C_{02}}=0$. According to the analytical solution for the equal-time second-order correlation function, if ${C_{20}}=0$ and ${C_{02}}=0$ then $g_a^{(2)}(0) \to 0$ and $g_b^{(2)}(0) \to 0$ generating the complete photon blockade effect. Thus, we obtain the optimal parameter conditions for the simultaneous nonreciprocal unconventional photon blockade when the laser drives the cavity from the left side. According to Eqs.~(\ref{afenzi0}) and~(\ref{bfenzi0}) (more detail see Appendix \ref{A}), we are able to derive
\begin{widetext}
\begin{eqnarray}
{\zeta _{{\rm{opt}}}} &=& \frac{\sqrt{F} \left( 4 \Delta^2 + \kappa^2 \right)^{\frac{1}{4}} \left( 16 \left( J^2 - \Delta^2 + \Delta_{SF}^2 \right)^2 +C \right)^{\frac{1}{4}}}{2J}
,\label{zetaopt}\\
{\varphi_{{\rm{opt}}}} &=& -\frac{1}{2} i \log\left[-\frac{\left( 2 \Delta - i \kappa \right) \left( 4 \left( J^2 - \Delta^2 + \Delta_{SF}^2 \right) + 4 i \Delta \kappa + \kappa^2 \right)}{\sqrt{4 \Delta^2 + \kappa^2} \sqrt{16 \left( J^2 - \Delta^2 + \Delta_{SF}^2 \right)^2 + C}}\right]
,\label{phiopt}\\
{K _{{\rm{opt}}}} &=& \frac{F \sqrt{4 \left( J^2 + 2 \Delta \left( \Delta - \Delta_{SF} \right) \right)^2 + D}}{2 J^2}
,\label{Kopt}\\
{\theta _{{\rm{opt}}}} &=& -i \log\left[\frac{2 J^2 + \left( 2 \Delta + i \kappa \right) \left( 2 \Delta - 2 \Delta_{SF} + i \kappa \right)}{\sqrt{4 \left( J^2 + 2 \Delta \left( \Delta - \Delta_{SF} \right) \right)^2 +D}}\right]
.\label{thetaopt}
\end{eqnarray}
\end{widetext}
When $\Delta$, $J$, $\Delta _{SF}$, $F$, and $\kappa$ are determined, the optimal parameters $\zeta _{{\rm{opt}}}$, $\phi_{{\rm{opt}}}$, $K _{{\rm{opt}}}$, and $\theta _{{\rm{opt}}}$ can be obtained from Eqs.~(\ref{zetaopt}) - (\ref{thetaopt}).
\begin{figure}[t]
\centerline{
\includegraphics[width=8.5cm, height=7.02cm, clip]{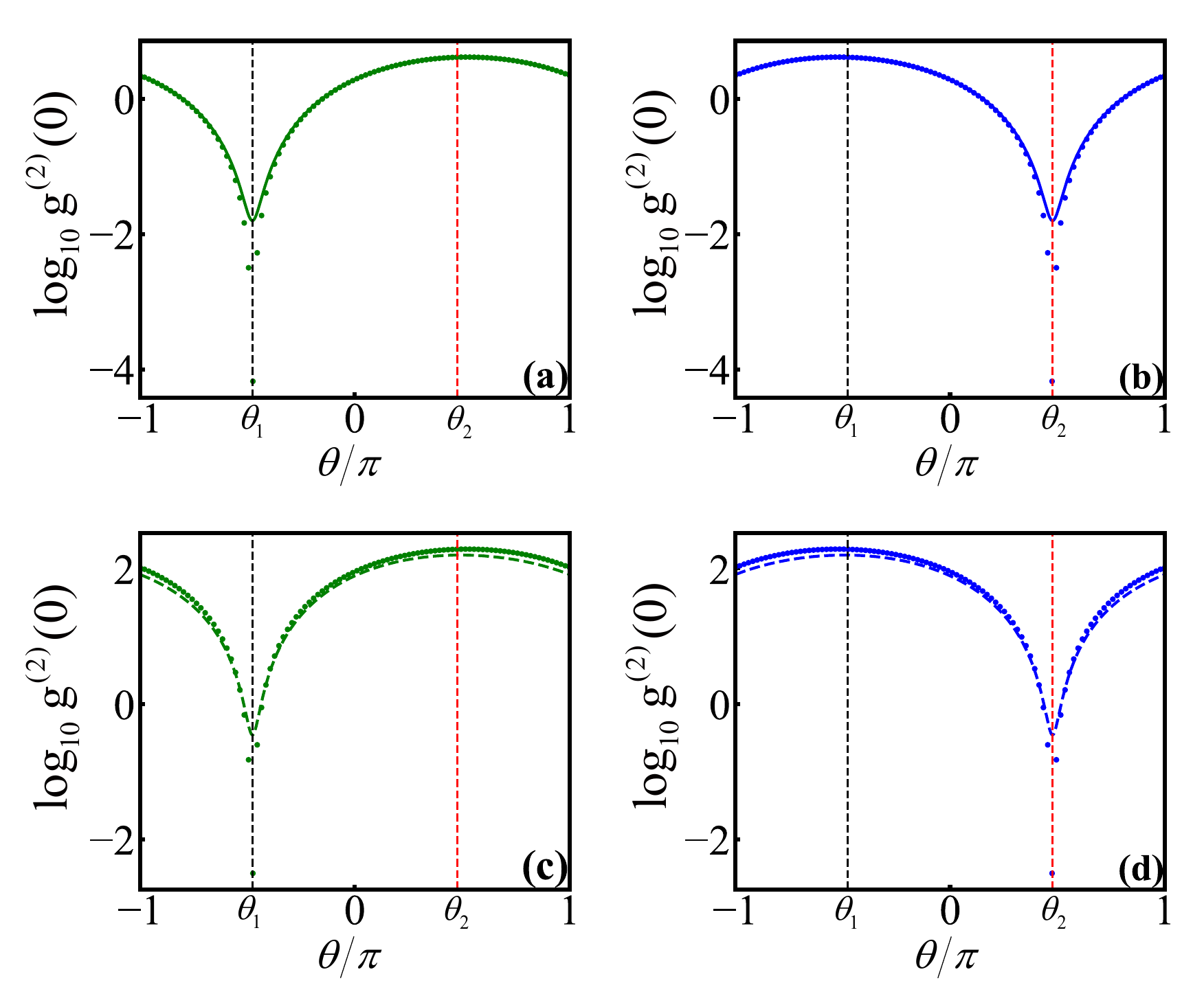}}
\vspace{-0.1cm}
\caption{(Color online) Second-order correlation function on a logarithmic scale ${\log _{10}}g_a^{(2)}(0)$ and ${\log _{10}}g_b^{(2)}(0)$ as a function of $\theta /\pi $ for ${\Delta _{SF}}/\kappa  > 0$ (in green) and ${\Delta _{SF}}/\kappa  < 0$ (in blue), respectively. The solid line (mode $a$) and dashed line (mode $b$) correspond to numerical simulations given by Eq.~(\ref{szg20}), while the dotted line corresponds to analytical solutions of Eq.~(\ref{jxg20}). The parameters are $\Delta/\kappa=0$, $\Delta_{SF}/\kappa=\pm7$, $J/\kappa=1$, $\zeta/\kappa=\zeta _{{\rm{opt}}}$, $\varphi=\varphi_{{\rm{opt}}}$, $K/\kappa=K _{{\rm{opt}}}$ and $F/\kappa=0.001$.}\label{jxsz}
\end{figure}

To verify the above analysis, we investigate the influences of the phase of two-photon driving in cavity $a$ on the photon antibunching characteristics. Figure~\ref{jxsz} displays the equal-time second-order correlation function $g_a^{(2)}(0)$ and $g_b^{(2)}(0)$ as a function of the phase $\theta /\pi $ for $J/\kappa  = 1$ and ${\Delta _{SF}}/\kappa  = 7$ in Fig.~\ref{jxsz}(a) - (d). Figs.~\ref{jxsz}(a) and~\ref{jxsz}(c) depict $g_a^{(2)}(0)$ and $g_b^{(2)}(0)$ for the case where ${\Delta _{SF}}/\kappa  > 0$ and Figs.~\ref{jxsz}(b) and~\ref{jxsz}(d) correspond to the case where ${\Delta _{SF}}/\kappa  < 0$, with $\Delta/\kappa = 0$ and $F/\kappa=0.001$. When the probe field inputs from left, i.e.,  ${\Delta _{SF}}/\kappa  > 0$, it is obvious that the two optical modes simultaneously exhibit a strong antibunching effect at optimal phase $\theta/\pi=-0.477$ (labeled by a vertical dashed line). The origin of strong photon antibunching  in two modes is the destructive quantum interference between direct and indirect paths in two-photon excitation. When $\Delta=0$, we can obtain
\begin{widetext}
\begin{eqnarray}
\zeta _{{\rm{opt}}} &=& \frac{{\sqrt F {{\left( {{\kappa ^2}} \right)}^{1/4}}{{\left( {16{{\left( {{J^2} + \Delta _{SF}^2} \right)}^2} + 8\left( {{J^2} + \Delta _{SF}^2} \right){\kappa ^2} + {\kappa ^4}} \right)}^{1/4}}}}{{2J}},\label{zetaopt0}\\
\varphi_{{\rm{opt}}} &=&  - \frac{1}{2}{i}\log \left[ {\frac{{{i}\kappa \left( {4\left( {{J^2} + \Delta _{SF}^2} \right) + {\kappa ^2}} \right)}}{{\sqrt {{\kappa ^2}} \sqrt {16{{\left( {{J^2} + \Delta _{SF}^2} \right)}^2} + 8\left( {{J^2} + \Delta _{SF}^2} \right){\kappa ^2} + {\kappa ^4}} }}} \right] ,\label{phiopt0}\\
K _{{\rm{opt}}} &=& \frac{{F\sqrt {4{J^4} + 4\left( { - {J^2} + \Delta _{SF}^2} \right){\kappa ^2} + {\kappa ^4}} }}{{2{J^2}}},\label{Kopt0}\\
\theta _{{\rm{opt}}} &=& - {i}\log \left[ {\frac{{2{J^2} + {i}\left( { - 2{\Delta _{SF}} + {i}\kappa } \right)\kappa }}{{\sqrt {4{J^4} + 4\left( { - {J^2} + \Delta _{SF}^2} \right){\kappa ^2} + {\kappa ^4}} }}} \right].\label{thetaopt0}
\end{eqnarray}
\end{widetext}

We find that the sign of $\Delta_{SF}$ has no influences on $\zeta _{{\rm{opt}}}$, $\varphi_{{\rm{opt}}}$ and $K _{{\rm{opt}}}$, but it does affect $\theta _{{\rm{opt}}}$. Thus, when we take $-\Delta_{SF}$, it can be observed that simultaneous nonreciprocal unconventional photon blockade also occurs at $\theta/\pi=0.477$. The analytical (dotted line) and numerical results (solid line and dashed line) are validated by our above analyses, both having a similar trend in the figure and the same optimal points acquiring the minimum
value of $g_a^{(2)}(0)$ and $g_b^{(2)}(0)$.
\begin{figure}[t]
\centerline{
\includegraphics[width=8.5cm, height=6.02cm, clip]{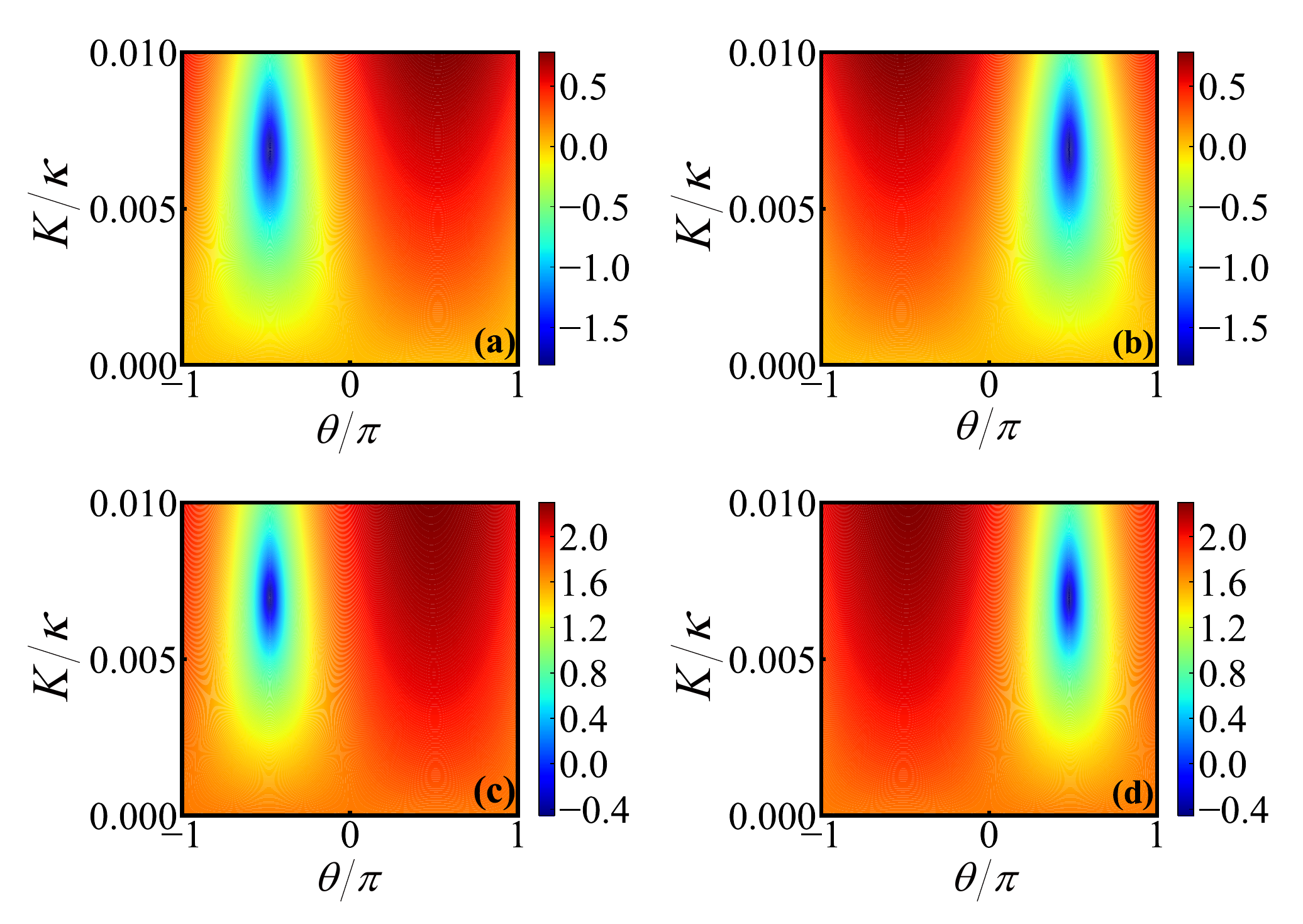}}
\vspace{-0.1cm}
\caption{(Color online) Second-order correlation function on a logarithmic scale (a) ${\log _{10}}g_a^{(2)}(0)$ and (c) ${\log _{10}}g_b^{(2)}(0)$ for $\Delta_{SF}>0$, and (b) ${\log _{10}}g_a^{(2)}(0)$ and (d) ${\log _{10}}g_b^{(2)}(0)$ for $\Delta_{SF}<0$ as functions of phase $\theta/\kappa$ and gain $K/\kappa$. The parameters are the same as in Fig.~\ref{jxsz}.}\label{SNPB3}
\end{figure}

Figures.~\ref{SNPB3}(a) and~\ref{SNPB3}(c) show the logarithmic values of the second-order correlation functions $g_a^{(2)}(0)$ and $g_b^{(2)}(0)$, respectively, as functions of the phase $\theta/\pi$ and the parametric gain $K/\kappa$ for $\Delta_{SF}>0$. The dark blue areas correspond to photon antibunching with sub-Poissonian statistics $[g_a^{(2)}(0)<1$ or $g_b^{(2)}(0)<1]$. At $\theta=\theta _{{\rm{opt}}}$ and $K=K _{{\rm{opt}}}$, photon blockade can be observed simultaneously in both cavities. When the probe field input is from port right, i.e. $\Delta_{SF}<0$, simultaneous photon antibunching appears in two cavities at $\theta=-\theta _{{\rm{opt}}}$, as shown in Figs.~\ref{SNPB3}(b) and~\ref{SNPB3}(d). The photon statistics for the two different end inputs are symmetric about the line $\theta/\pi=0$.
\begin{figure}[t]
\centerline{
\includegraphics[width=8.5cm, height=3.35cm, clip]{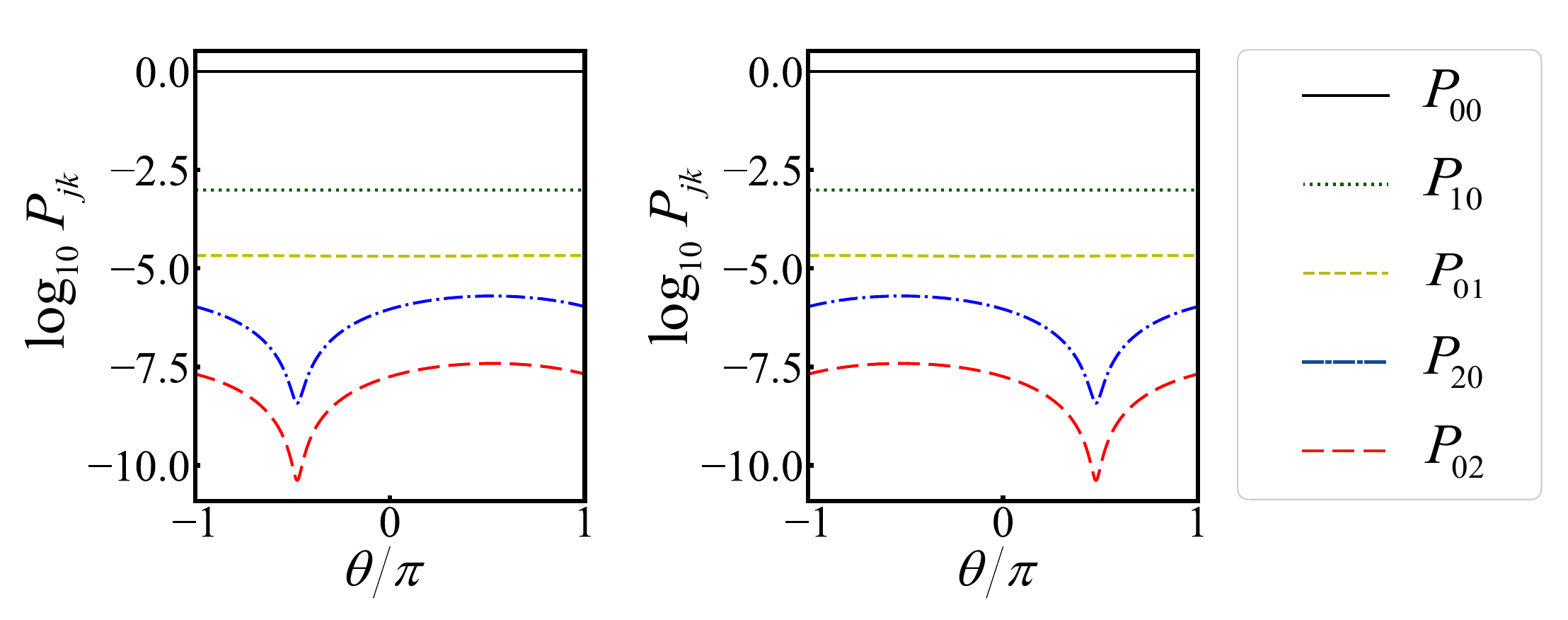}}
\vspace{-0.1cm}
\caption{(Color online) State occupations on a logarithmic scale ${\log _{10}}{P_{mn}}$ as a function of $\theta/\kappa$ for (a) $\Delta_{SF}>0$ and (b) $\Delta_{SF}<0$. The parameters are the same as in Fig.~\ref{jxsz}. }\label{Pjk}
\end{figure}

To show this more clearly, we plot the state occupations ${P_{mn}} = \left| {\left\langle {{mn}}\mathrel{\left | {\vphantom {{mn} \varphi }}\right. \kern-\nulldelimiterspace}
{\psi } \right\rangle } \right|$ $(m,n=0,1,2)$ for $\Delta_{SF}>0$ and $\Delta_{SF}<0$ in Figs.~\ref{Pjk}(a) and~\ref{Pjk}(b), respectively. It can be seen that at
$\theta=\theta _{{\rm{opt}}}$, states $\left| {20} \right\rangle $ and $\left| {02} \right\rangle $ show a significant decrease. Notably, the evolution of ${P_{20}}$ and ${P_{02}}$ closely mirrors the logarithmic trends of ${\log _{10}}g_a^2(0)$ and ${\log _{10}}g_b^2(0)$ in Fig.~\ref{jxsz},  respectively. This correlation confirms that the observed strong photon antibunching arises from quantum destructive interference between distinct transition pathways populating the states $\left| {20} \right\rangle $ and $\left| {02} \right\rangle $.
 \begin{figure}[t]
\centerline{
\includegraphics[width=8.5cm, height=2.43cm, clip]{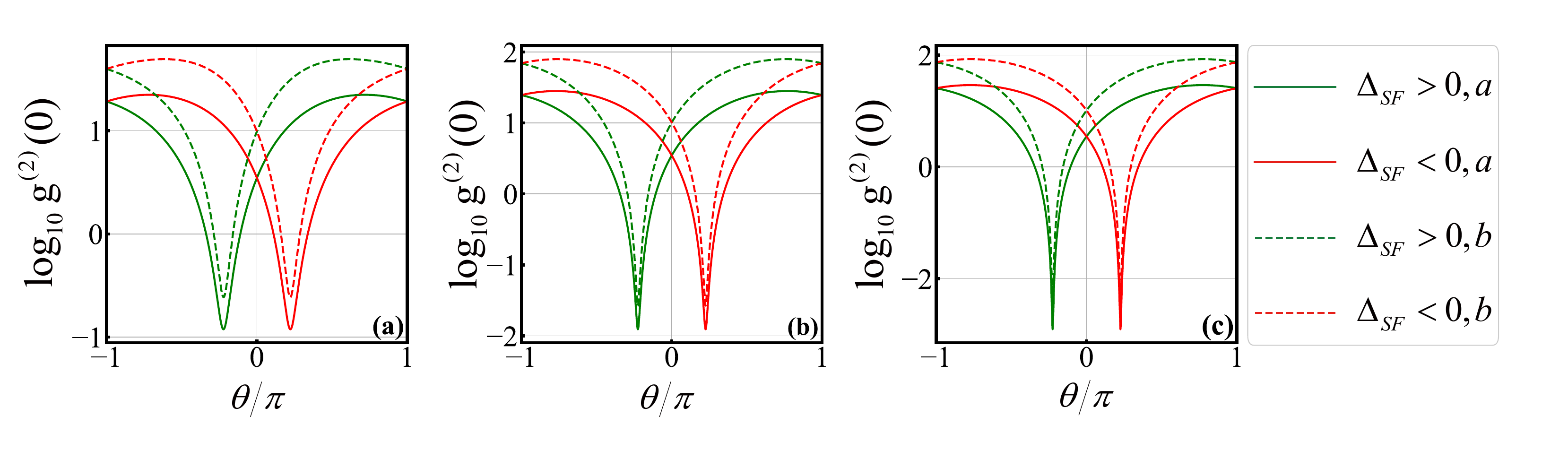}}
\vspace{-0.1cm}
\caption{(Color online) Second-order correlation function on a logarithmic
scale ${\log _{10}}g_a^{(2)}(0)$ and ${\log _{10}}g_b^{(2)}(0)$ as functions of phase $\theta/\kappa$ for different nonlinear gains, where (a)$F/\kappa=0.01$, (b) $F/\kappa=0.001$, (c) $F/\kappa=0.0001$. The parameters chosen are $\Delta/\kappa=0$, $\Delta_{SF}/\kappa=\pm3$, $J/\kappa=2$, $\zeta/\kappa=\zeta _{{\rm{opt}}}$, $\varphi=\varphi_{{\rm{opt}}}$, $K/\kappa=K _{{\rm{opt}}}$.}\label{diffF}
\end{figure}

In Fig.~\ref{diffF}, we show $g_a^{(2)}(0)$ and $g_b^{(2)}(0)$ as a function of $\theta /\pi$ with different moderate values of $F/\kappa$, with $J /\kappa=2$ and $\Delta_{SF}/\kappa=3$. In the figure, the green and red colors represent the cases of $\Delta_{SF}/\kappa>0$ and $\Delta_{SF}/\kappa<0$, respectively. The solid lines correspond to mode $a$, while mode $b$ is represented by the dashed lines. From  Eq.~(\ref{thetaopt0}), we can see that $\theta _{{\rm{opt}}}$ is independent of $F$ and $\theta _{{\rm{opt}}}$ remains equal to $-0.226\pi$. In Figs.~\ref{diffF}(a) - (c),  where $F$ is set to $0.01$, $0.001$, and $0.0001$, respectively, we can observe that at $\theta/\pi=-0.226$, both mode $a$ and mode $b$ simultaneously exhibit unconventional photon blockade. At this point, the correlation function associated with the case of $\Delta_{SF}/\kappa<0$ is greater than $0$.
\begin{figure}[t]
\centerline{
\includegraphics[width=8.5cm, height=2.41cm, clip]{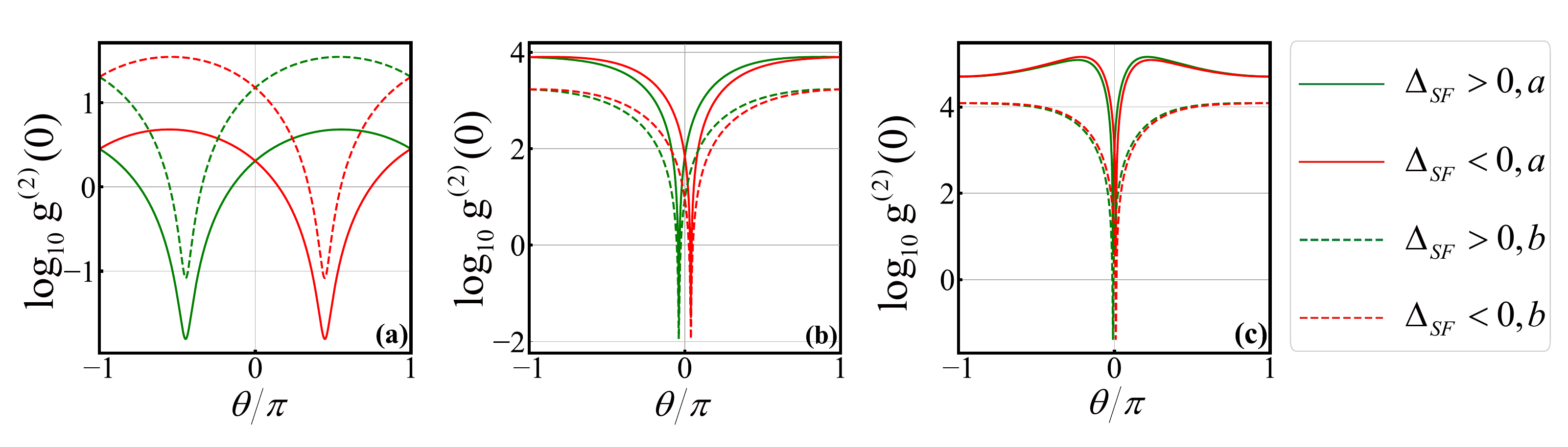}}
\vspace{-0.1cm}
\caption{(Color online) Second-order correlation function on a logarithmic
scale ${\log _{10}}g_a^{(2)}(0)$ and ${\log _{10}}g_b^{(2)}(0)$ as functions of phase $\theta/\kappa$ for different coupling strength $J$, where (a)$J/\kappa=1$, (b) $J/\kappa=5$, (c) $J/\kappa=10$. The parameters are $\Delta/\kappa=0$, $\Delta_{SF}/\kappa=\pm3$, $F/\kappa=0.001$, $\zeta/\kappa=\zeta _{{\rm{opt}}}$, $\varphi=\varphi_{{\rm{opt}}}$, $K/\kappa=K _{{\rm{opt}}}$.}\label{diffJ}
\end{figure}
\begin{figure}[t]
\centerline{
\includegraphics[width=8.5cm, height=6.5cm, clip]{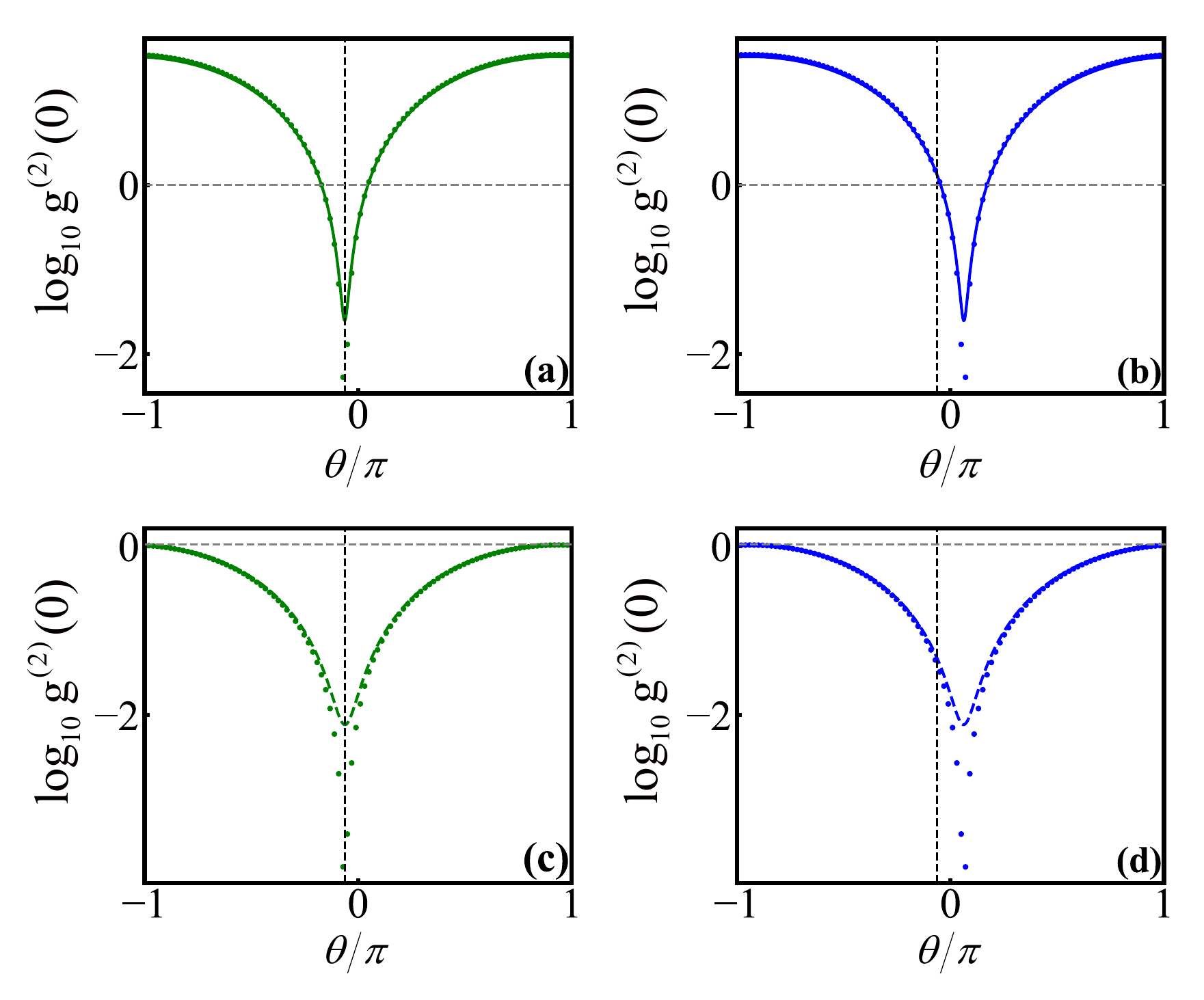}}
\vspace{-0.1cm}
\caption{(Color online) Second-order correlation functions are plotted on a logarithmic scale, where ${\log _{10}}g_a^{(2)}(0)$ and ${\log _{10}}g_b^{(2)}(0)$ are displayed as functions of the DOPA phase $\theta /\pi $. The solid line (mode $a$) and dashed line (mode $b$) correspond to numerical simulations given by Eq.~(\ref{szg20}), while the dotted line corresponds to analytical solutions of Eq.~(\ref{jxg20}). The selected parameters are $\Delta_{SF}/\kappa=\pm0.1$, with other parameters identical to those in Fig.~\ref{jxsz}.
}\label{0.1DeltaF}
\end{figure}
\begin{figure}[t]
\centerline{
\includegraphics[width=8.5cm, height=6cm, clip]{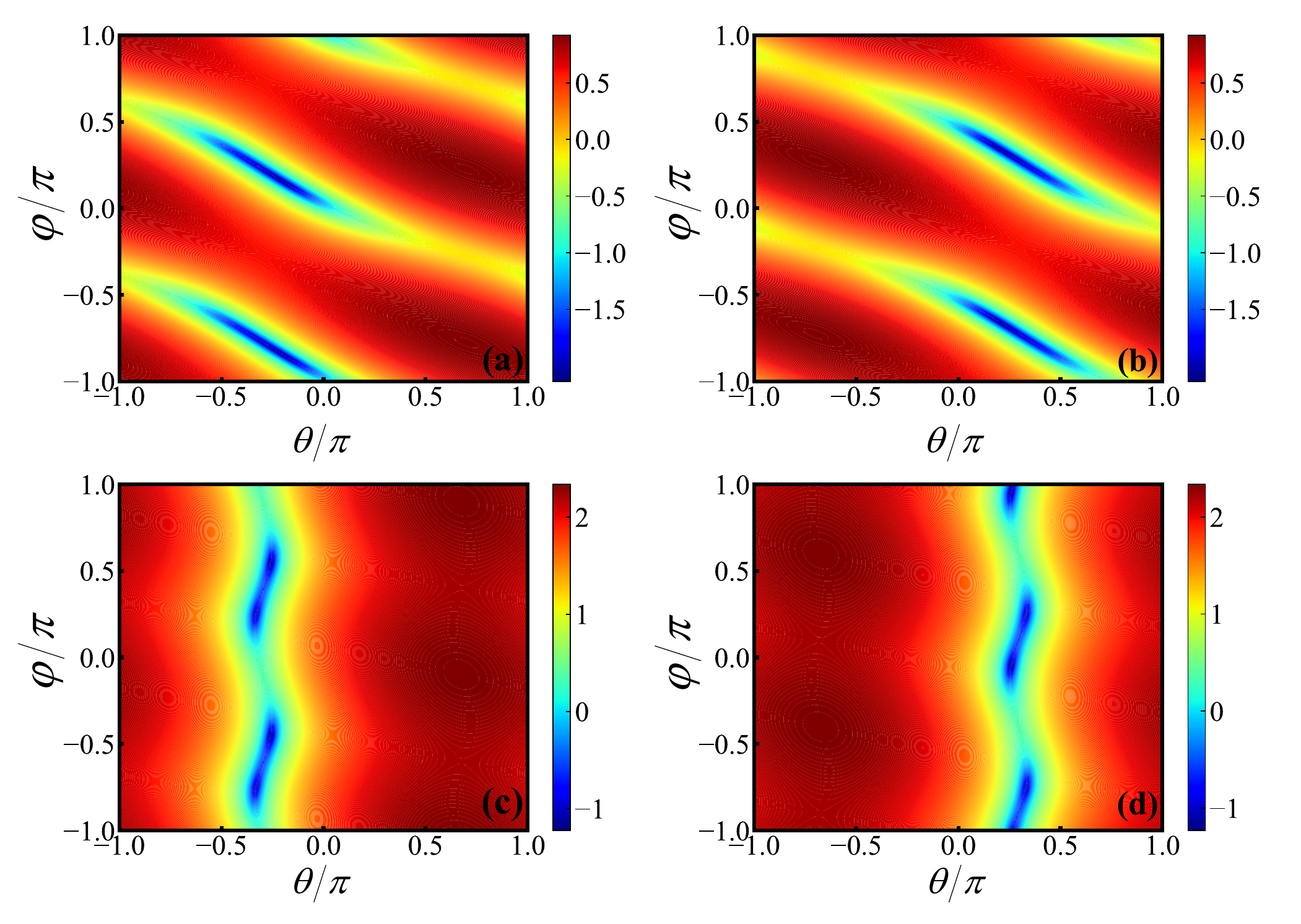}}
\vspace{-0.1cm}
\caption{(Color online) Second-order correlation function on a logarithmic scale (a) ${\log _{10}}g_a^{(2)}(0)$ and (c) ${\log _{10}}g_b^{(2)}(0)$ for $\Delta_{SF}>0$, and (b) ${\log _{10}}g_a^{(2)}(0)$ and (d) ${\log _{10}}g_b^{(2)}(0)$ for $\Delta_{SF}<0$ as functions of phase $\theta/\kappa$ and phase $\varphi/\kappa$. The parameters are the same as in Fig.~\ref{jxsz}.}\label{thetaphi}
\end{figure}
Next, we plotted Fig.~\ref{diffJ}, which differs from Fig.~\ref{diffF} in that $F$ is fixed at $0.001$ while $J$ takes different values. Figs.~\ref{diffJ}(a) - (c) correspond to $J/\kappa = 1$, $J/\kappa = 5$, and $J/\kappa = 10$, respectively. The figure shows that the correlation function exhibits pronounced oscillatory behavior and is symmetric about
$\theta=0$. As $J$ increases, stronger peaks or dips appear near $\theta/\pi=0$, and the difference between the solid and dashed lines becomes smaller, indicating that the effect of the sign of $\Delta_{SF}$ on the correlation function gradually weakens, suggesting that nonreciprocity is gradually disappearing.
\begin{figure}[t]
\centerline{
\includegraphics[width=4.5cm, height=3.09cm, clip]{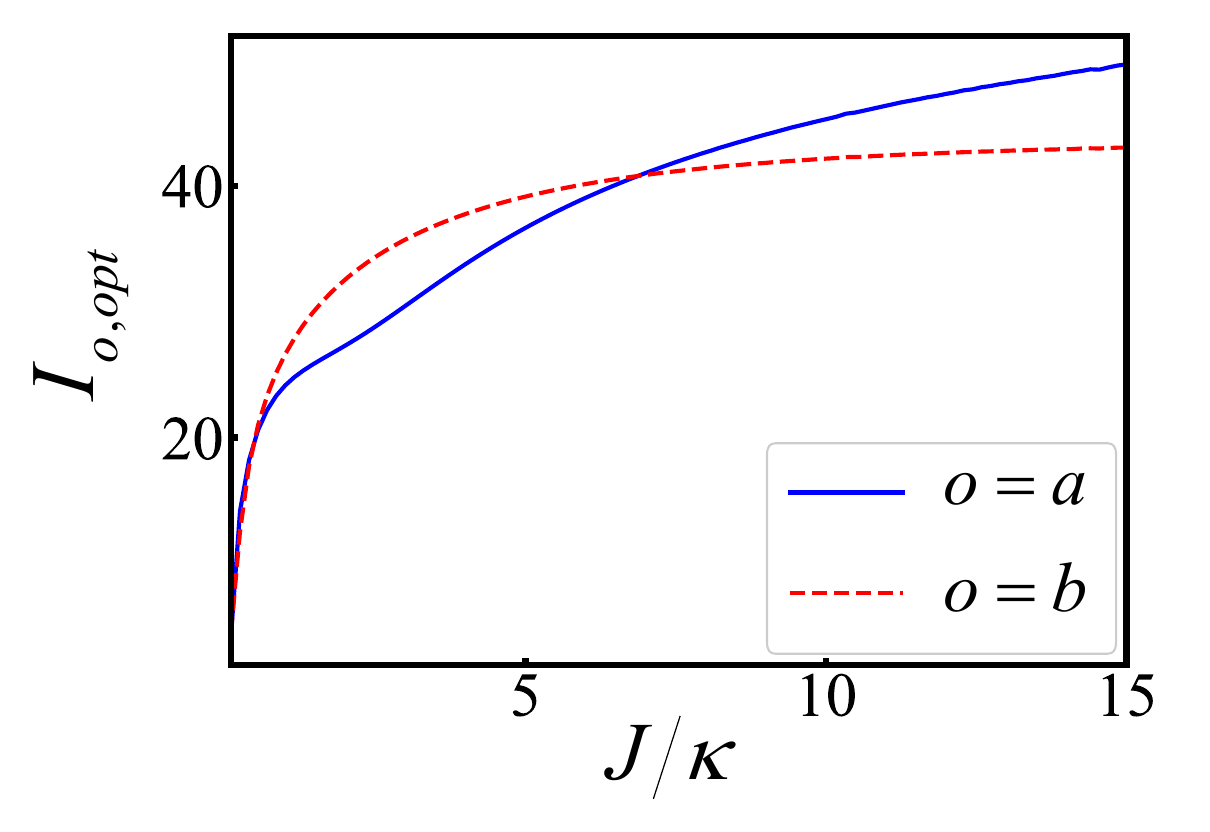}}
\vspace{-0.1cm}
\caption{(Color online) The nonreciprocal ratio ${I_{o,{\rm{opt}}}}$ as a function of coupling strength $J/\kappa$. The parameters are $\zeta/\kappa=\zeta _{{\rm{opt}}}$, $\varphi=\varphi_{{\rm{opt}}}$, $K/\kappa=K _{{\rm{opt}}}$ and $\theta/\kappa=\theta _{{\rm{opt}}}$. The parameters are the same as in Fig.~\ref{jxsz}.}\label{figratio}
\end{figure}
\begin{figure}[t]
\centerline{
\includegraphics[width=4.5cm, height=3.64cm, clip]{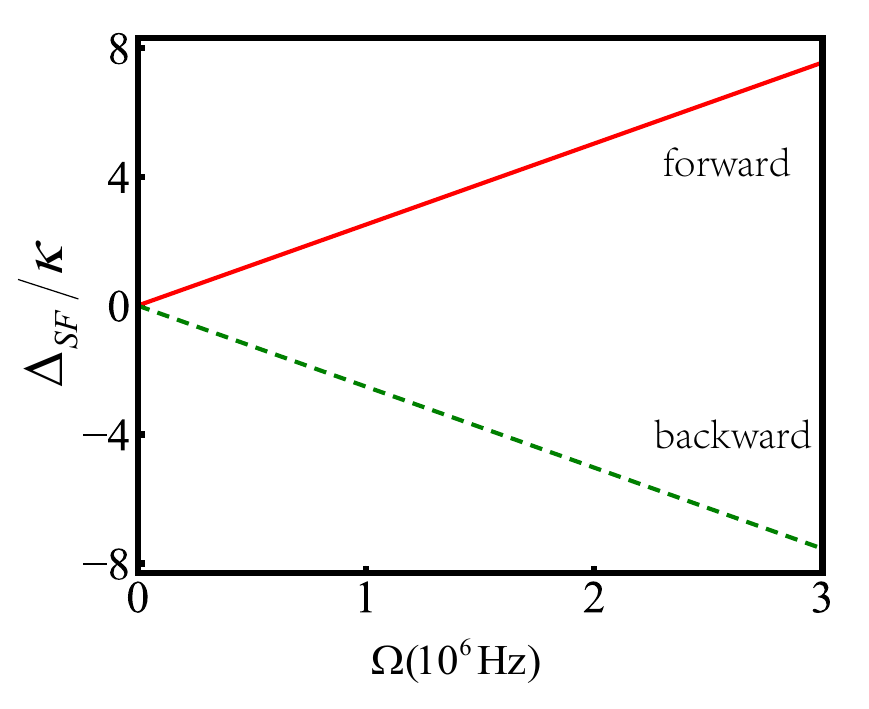}}
\vspace{-0.1cm}
\caption{(Color online) The ratio of Fizeau drag to decay $\Delta_{SF}/\kappa$ as a function of angular velocity $\Omega$ for $\Delta_{SF}/\kappa>0$ (red line) and $\Delta_{SF}/\kappa<0$ (green line). Increasing the angular velocity results in a linear opposite frequency shift for the counterpropagating modes.}\label{DeltaFOmega}
\end{figure}

To ensure data integrity, we also analyzed the scenario with varying ${\Delta _{SF}}/\kappa$ values, as illustrated in Fig.~\ref{0.1DeltaF}. Compared with Fig.~\ref{0.1DeltaF}, the optimized parameter $\theta_{{\rm{opt}}}$ shifts from $-0.477\pi$ to $-0.06\pi$. The nonreciprocity between Port 1 and Port 2 gradually diminishes. For mode $a$, the second-order correlation function remains below 1 when driven from Port 1 but exceeds 1 when driven from Port 2, preserving the nonreciprocal photon blockade characteristic. However, for mode $b$, the second-order correlation functions under both driving directions at the optimized phase are below 1, indicating photon blockade regardless of the driving direction. This demonstrates that by tuning ${\Delta _{SF}}/\kappa$, the system can achieve either simultaneous nonreciprocal unconventional photon blockade in both modes or nonreciprocal blockade in one mode (e.g., mode $a$) while the other (mode $b$) remains fully blocked at the optimized phase. Such tunability holds broad potential for future quantum devices.

Next, we plotted the second-order correlation function on a logarithmic scale as a function of $\theta/\kappa$ and phase $\varphi/\kappa$. Figs.~\ref{thetaphi}(a) and~\ref{thetaphi}(c) represent the case when $\Delta_{SF}/\kappa>0$, while figs.~\ref{thetaphi}(b) and~\ref{thetaphi}(d) represent the case when $\Delta_{SF}/\kappa<0$. Figs.~\ref{thetaphi}(a) and~\ref{thetaphi}(b) correspond to mode a, while figs.~\ref{thetaphi}(c) and~\ref{thetaphi}(d) represent mode b. From the figures, we can see that when $\zeta=\zeta _{{\rm{opt}}}$,  $K=K _{{\rm{opt}}}$, the plots change periodically with $\theta/\kappa$ and phase $\varphi/\kappa$. Due to the different signs of  $\Delta_{SF}/\kappa$, the plots are not the same. By adjusting the values of $\theta/\kappa$ and phase $\varphi/\kappa$, we can obtain the range where mode $a$ and mode $b$ are both blocked when $\Delta_{SF}/\kappa>0$. Within this range, when $\Delta_{SF}/\kappa<0$ no blockade occurs. It can be seen that nonreciprocal photon blockade is not limited to a specific value, but can be realized within a certain range. Additionally, the direction of the blockade can be changed by adjusting the value of $\theta/\kappa$.

To quantitatively characterize the degree of nonreciprocal photon blockade under optimal conditions, we define a nonreciprocity ratio
\begin{equation}
\begin{aligned}
{I_o} =  - 10 {\rm{lo}}{{\rm{g}}_{10}}[\frac{{g_{o,{\Delta _{SF}} > 0}^{(2)}(0)}}{{g_{o,{\Delta _{SF}} < 0}^{(2)}(0)}}].\label{ratio}
\end{aligned}
\end{equation}
for mode o. This ratio quantifies the contrast between photon correlation functions generated from two distinct input configurations. The formulation of this ratio to quantify nonreciprocal photon blockade stems from the stark divergence in photon statistics observed at optimal parameters, where one input configuration yields a deeply suppressed second-order correlation $g_o^{(2)}(0) \ll 1$, while the opposite input drives it to near-classical levels $g_o^{(2)}(0)\sim1$. From Fig.~\ref{figratio}, we see that a larger coupling strength g generates a better nonreciprocity for both modes, where we use ${I_{o,{\rm{opt}}}}$ to represent the nonreciprocal ratio under the optimal parameters. This is
because that a large $J$ involved in key excitation paths can enhance the quantum destructive interference, leading to a smaller $g_a^{(2)}(0)$ and $g_b^{(2)}(0)$.

In our calculations, for ensuring the stability of the system, we use the experimentally feasible values \cite{Vahala8392003,Spillane0138172005,Aspelmeyer13912014,Peng3942014}: $n$=1.4, $R$=1.1mm, $\lambda$=1.55$\mu$m, $\omega_{c}$=2$\pi$THz, and the vaule we use for $\kappa$ is 2$\pi$MHz. To achieve gigahertz regime, a levitated optomechanical device can be utilized, which allow us to explore a wider range of parameter possibilities. From Fig.~\ref{DeltaFOmega}, we can observe that when $\Omega$  reaches 3MHz, and $\Delta_{SF}$/$\kappa$ exceeds 6, which aligns with the $\Delta_{SF}$ values used in the article.
\begin{figure}[t]
\centerline{
\includegraphics[width=8.5cm, height=2.75cm, clip]{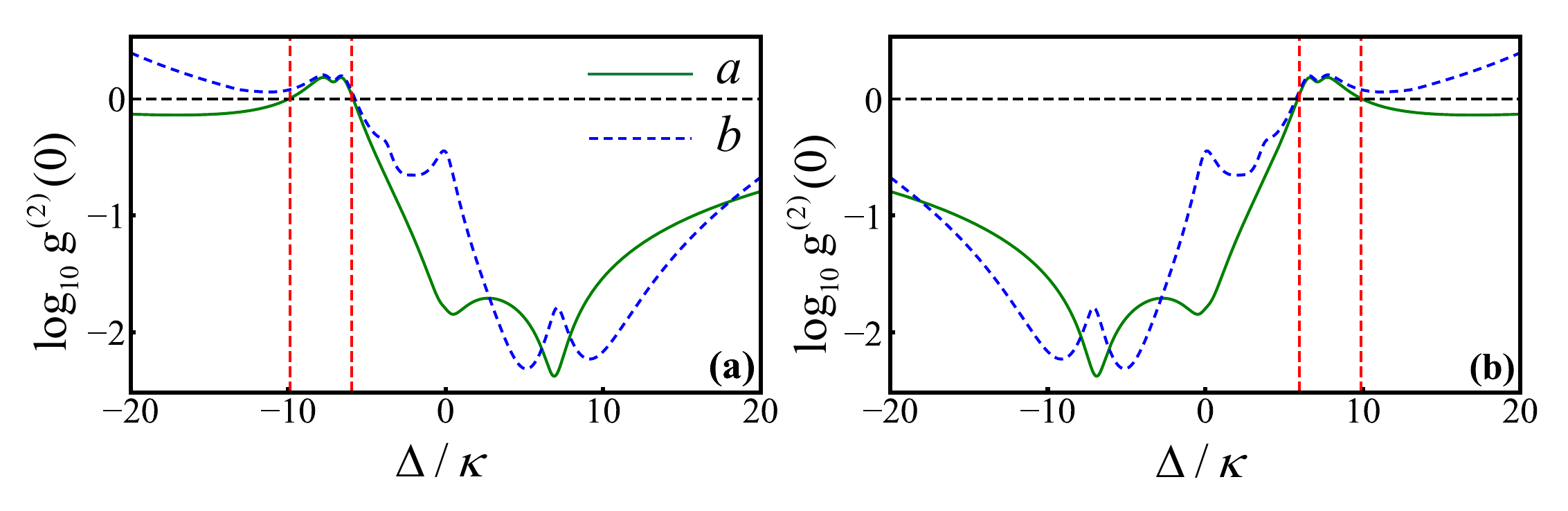}}
\vspace{-0.1cm}
\caption{(Color online) The second-order correlation functions ${\log _{10}}g_o^{(2)}(0) (o=a, b)$ are plotted on a logarithmic scale as functions of the normalized detuning $\Delta/\kappa$. Here, mode $a$ is represented by a green solid line, and mode $b$ by a blue dashed line. A black dashed line marks the position ${\log _{10}}g_o^{(2)}(0)=0$, while two red dashed lines indicate the range where the second-order correlation functions of both modes exceed 1. The selected parameter is $\theta _{{\rm{opt}}}$, with other parameters identical to those in Fig.~\ref{jxsz}. }\label{Deltag20}
\end{figure}
\begin{figure}[t]
\centerline{
\includegraphics[width=8.5cm, height=5.4cm, clip]{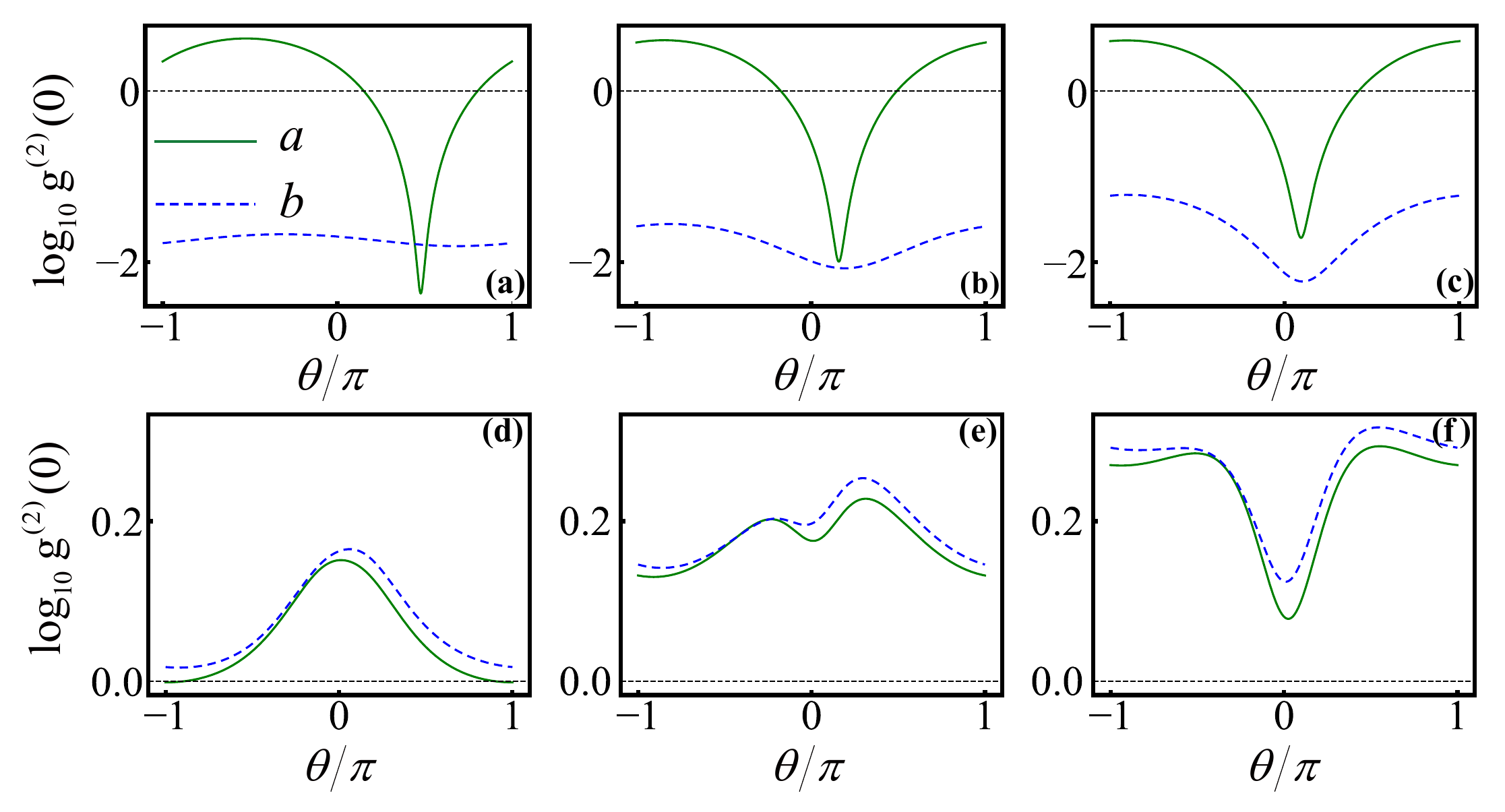}}
\vspace{-0.1cm}
\caption{(Color online) The second-order correlation functions ${\log _{10}}g_o^{(2)}(0) (o=a, b)$ are plotted on a logarithmic scale as functions of the normalized detuning
$\theta/\kappa$. Mode $a$ is represented by a green solid line, and mode $b$ by a blue dashed line. The first row corresponds to  $\Delta_{SF}/\kappa>0$, while the second row represents the case of $\Delta_{SF}/\kappa<0$. Each column corresponds to the detuning values $\Delta/\kappa=7,8,9$, respectively. The selected parameters are identical to those in Fig.~\ref{jxsz}.
}\label{thetag20Delta}
\end{figure}

The previous section introduce the phenomenon of simultaneous unconventional photon blockade when $\Delta = 0$. We now discuss the case of simultaneous nonreciprocal unconventional photon blockade when $\Delta \neq 0$. For $\Delta \neq 0$, Fig.~\ref{Deltag20} plots the second-order correlation functions ${\log _{10}}g_a^{(2)}(0)$ and ${\log _{10}}g_b^{(2)}(0)$ as functions of the normalized detuning $\Delta/\kappa$. Fig.~\ref{Deltag20}(a) corresponds to $\Delta_{SF}/\kappa>0$, while Fig.~\ref{Deltag20}(b) represents $\Delta_{SF}<0$. From the figures, it can be observed that the regions marked by red dashed lines exhibit simultaneous nonreciprocal unconventional photon blockade. In Fig.~\ref{Deltag20}(a), when $\Delta/\kappa$ lies between $-9.874$ and $-5.964$, the second-order correlation functions of both modes are greater than 1. However, within the same detuning range in Fig.~\ref{Deltag20}(b), the values are less than 1, allowing us to confirm the occurrence of simultaneous nonreciprocal unconventional photon blockade. Similarly, when photons are input backward, simultaneous nonreciprocal unconventional photon blockade is also achieved within the detuning range of $5.964$ to $9.874$ marked by the red line in Fig.~\ref{Deltag20}(b). These two ranges are symmetric about $\Delta/\kappa=0$, enabling simultaneous nonreciprocal unconventional photon blockade in both $\Delta_{SF}/\kappa>0$ and $\Delta_{SF}/\kappa<0$ regions.

Next, we select detunings $\Delta/\kappa$=7,8,9 to plot ${\log _{10}}g_a^{(2)}(0)$ and ${\log _{10}}g_b^{(2)}(0)$ as functions of the phase $\theta/\pi$. Figs.~\ref{thetag20Delta}(a) - (c) correspond to $\Delta_{SF}/\kappa>0$, while Figs.~\ref{thetag20Delta}(d) - (f) represent $\Delta_{SF}/\kappa<0$. Each vertical column corresponds to a specific detuning value. From the figures, it can be observed that in every plot with $\Delta_{SF}/\kappa<0$, there are regions where the second-order correlation functions of both modes are simultaneously less than 1. Within the same phase ranges, for $\Delta_{SF}/\kappa>0$, the second-order correlation functions of both modes exceed 1, indicating simultaneous nonreciprocal unconventional photon blockade. Notably, in Figs.~\ref{thetag20Delta}(a) - (c), there exist cases where the second-order correlation function of mode $a$ is greater than 1, while that of mode $b$ is less than 1. This demonstrates that by adjusting the phase $\theta/\pi$, it is possible to realize a scenario where one mode exhibits reciprocity (blocked in both directions) while the other remains nonreciprocal.

\section{CONCLUSIONS}\label{Sec5}
In summary, we propose a scheme to achieve simultaneous nonreciprocal unconventional photon blockade in two coupled rotating cavities with DOPA. When light enters from two opposite ports, it induces Sagnac-Fizeau shifts of equal magnitude but opposite signs in the cavities rotating at a fixed angular velocity, thereby generating nonreciprocal phenomena. Our results demonstrate that under optimized parameter conditions, both modes $a$ and $b$ exhibit photon blockade simultaneously under forward driving. However, for backward driving at these parameters, the opposite driving direction leads to reversed Sagnac-Fizeau shifts, preventing photon blockade. This behavior primarily depends on whether quantum destructive interference occurs in the two-photon excitation process. For input from one direction, quantum destructive interference enables simultaneous unconventional photon blockade in both cavities. Conversely, input from the opposite direction fails to induce complete quantum destructive interference, resulting in the disappearance of photon blockade. We specifically discuss cases with detuning $\Delta/\kappa = 0$ and $\Delta/\kappa \ne 0$. The results show that simultaneous nonreciprocal unconventional photon blockade can be achieved within a certain parameter range regardless of the presence of detuning. By adjusting the phase $\theta/\pi$, the direction of photon blockade can be reversed. In our system, when $\theta/\pi < 0$, simultaneous unconventional photon blockade occurs in both modes under forward driving, whereas $\theta/\pi > 0$ enables simultaneous blockade under backward driving. Notably, when adjusting $\Delta_{SF}/\kappa$ to 0.1, the system transitions from simultaneous nonreciprocal unconventional photon blockade in both modes at optimized phases to nonreciprocal photon blockade in mode $a$ alone, while mode $b$ exhibits photon blockade in both directions. This demonstrates that tuning $\Delta_{SF}/\kappa$  can modify the photon statistical properties of the system, realizing a controllable nonreciprocal platform.

\section*{ACKNOWLEDGMENTS}
This work was supported by Science and Technology Development Plan Project of Jilin Province (Grant No. 20250102007JC), National Natural Science Foundation of China under Grant No. 12274064, Scientific Research Project for Department of Education of Jilin Province under Grant No. JJKH20241410KJ.

\appendix
\section{DERIVATION OF OPTIMAL CONDITIONS}\label{A}
If ${C_{20}}=0$ and ${C_{02}}=0$ then $g_a^{(2)}(0) \to 0$ and $g_b^{(2)}(0) \to 0$ generating a complete photon blockade effect. So we need to set the numerators of $C_{20}$ and $C_{02}$ to zero, i.e,
\begin{small}
\begin{widetext}
\begin{eqnarray}
0 &=& \sqrt 2 {e^{ - i\theta }}(2{e^{i(\theta  + 2\varphi )}}{\zeta ^2}(2\Delta  - i\kappa ){{B}^2} + A(2{e^{i\theta }}F{J^2} - G(2{J^2} - 4{\Delta ^2} + 4\Delta {\Delta _{SF}} + 4i\kappa  - 2i{\Delta _{SF}}\kappa  + {\kappa ^2})),\label{afenzi0}\\
0 &=& \sqrt 2 {e^{ - i\theta }}(8{e^{i(\theta  + 2\varphi )}}{J^2}{\zeta ^2}(2\Delta  - i\kappa ) - A( - 2{J^2}K + {e^{i\theta }}F(2{J^2} - 4\Delta (\Delta  + {\Delta _{SF}}) + 2i(2\Delta  + {\Delta _{SF}})\kappa  + {\kappa ^2}))).\label{bfenzi0}
\end{eqnarray}
\end{widetext}
\end{small}
By decomposing the real and imaginary parts of the two molecules, setting them to zero, and solving the resulting system of equations, Eqs.~(\ref{zetaopt}) - (\ref{thetaopt}) can be obtained.


\begin{references}

\bibitem{Shields2052007}A. J. Shields, Semiconductor quantum light sources, Nat. Photon. \textbf{1}, 215 (2007).
\bibitem{Ghosh0136022019}S. Ghosh and T. C. H. Liew, Dynamical blockade in a singlemode bosonic system, Phys. Rev. Lett. \textbf{123}, 013602 (2019).
\bibitem{Huang21004302022}R. Huang, \c{S}. K. \"{O}zdemir, J. Q. Liao, F. Minganti, L. M. Kuang, F. Nori, and H. Jing, Exceptional photon blockade:Engineering photon blockade with chiral exceptional points, Laser Photon. Rev. \textbf{16}, 2100430 (2022).
\bibitem{Tang47052025}J. Tang, Y. Zuo, X. W. Xu, R. Huang, A. Miranowicz, F. Nori and H. Jing, Achieving Robust Single-Photon Blockade with a Single Nanotip, Nano. Lett. \textbf{25}, 4705-4712 (2025).
\bibitem{Zuo0437152022}Y. L. Zuo, R. Huang, L. M. Kuang, X. W. Xu, and H. Jing, Loss-induced suppression, revival, and switch of photon blockade, Phys. Rev. A \textbf{106}, 043715 (2022).
\bibitem{Liu0637052024}Z.-H. Liu and X.-W. Xu, Scaling enhancement of photon blockade in output fields, Phys. Rev. A \textbf{110}, 063705 (2024).
\bibitem{Zhang0637052024}G.-Y. Zhang, Z.-H. Liu, and X.-W. Xu, Optimizing dynamical blockade via a particle-swarm-optimization algorithm, Phys. Rev. A \textbf{110}, 023718 (2024).
\bibitem{Zou0537102020}F. Zou, X.-Y. Zhang, X.-W. Xu, J.-F. Huang, and J.-Q. Liao, Multiphoton blockade in the two-photon Jaynes-Cummings model, Phys. Rev. A \textbf{102}, 053710 (2020).
\bibitem{Xu0638532016}X.-W. Xu, A.-X. Chen, and Y.-x. Liu, Phonon blockade in a nanomechanical resonator resonantly coupled to a qubit, Phys. Rev. A \textbf{94}, 063853 (2016).
\bibitem{Guo0137052022}Y. T. Guo, F. Zou, J. F. Huang, and J. Q. Liao, Retrieval of photon blockade effect in the dispersive Jaynes-Cummings model, Phys. Rev. A \textbf{105}, 013705 (2022).
\bibitem{Li0437072024}H.-J. Li, L.-B. Fan, S. Ma, J.-Q. Liao, and C.-C. Shu, Exploring photon blockade in a two-photon Jaynes-Cummings model with atom and cavity drivings, Phys. Rev. A \textbf{110}, 043707 (2024).
\bibitem{Li24003742025}S. Li, N. Wang, and A.-D. Zhu, Sideband-Selective Single-Photon Blockade in Floquet-Modulated Jaynes-Cummings System, Adv. Quantum Technol. \textbf{8}, 2400374 (2025).
\bibitem{Lu0136022025}Z.-G. Lu, Y. Wu, and X.-Y. L¨¹, Chiral Interaction Induced Near-Perfect Photon Blockade, Phys. Rev. Lett. \textbf{134}, 013602 (2025).
\bibitem{Sun0437152023}J. Y. Sun and H. Z. Shen, Photon blockade in non-Hermitian optomechanical systems with nonreciprocal couplings, Phys. Rev. A \textbf{107}, 043715 (2023).
\bibitem{Gerace2812009} D. Gerace, H. E. T\"{u}reci, A. Imamo\={g}lu, V. Giovannetti, and R. Fazio, The quantum-optical Josephson interferometer, Nat. Phys. \textbf{5}, 281 (2009).
\bibitem{Chang8072007}D. E. Chang, A. S. S{\o}rensen, E. A. Demler, and M. D. Lukin, A single-photon transistor using nanoscale surface plasmons, Nat. Phys. \textbf{3}, 807 (2007).
\bibitem{Sayrin0410362015}C. Sayrin, C. Junge, R. Mitsch, B. Albrecht, D. O'Shea, P. Schneeweiss, J. Volz, and A. Rauschenbeutel, Nanophotonic optical isolator controlled by the internal state of cold atoms, Phys. Rev. X \textbf{5}, 041036 (2015).
\bibitem{Tang0438332019}L. Tang, J. Tang,W. Zhang, G. Lu, H. Zhang, Y. Zhang, K. Xia, and M. Xiao, On-chip chiral single-photon interface: Isolation and unidirectional emission, Phys. Rev. A \textbf{99}, 043833 (2019).
\bibitem{Zhao0638382020}C. Zhao, X. Li, S. Chao, R. Peng, C. Li, and L. Zhou, Simultaneous blockade of a photon, phonon, and magnon induced by a two-level atom, Phys. Rev. A \textbf{101}, 063838 (2020).
\bibitem{Feng123042023}L. J. Feng, L. Yan, and S. Q. Gong, Unconventional photon blockade induced by the self-Kerr and cross-Kerr nonlinearities, Front. Phys. \textbf{18}, 12304 (2023).
\bibitem{Bamba0218022011}M. Bamba, A. Imamo\={g}lu, I. Carusotto, and C. Ciuti, Origin of strong photon antibunching in weakly nonlinear photonic molecules, Phys. Rev. A \textbf{83}, 021802(R) (2011).
\bibitem{Birnbaum872005}K. M. Birnbaum, A. Boca, R. Miller, A. D. Boozer, T. E. Northup, and H. J. Kimble, Photon blockade in an optical cavity with one trapped atom, Nature (London) \textbf{436}, 87 (2005).
\bibitem{Reinhard932012}A. Reinhard, T. Volz, M. Winger, A. Badolato, K. J. Hennessy, E. L. Hu, and A. Imamo\v{g}lu, Strongly correlated photons on a chip, Nat. Photon. \textbf{6}, 93 (2012).
\bibitem{Muller2336012015}K. M\"{u}ller, A. Rundquist, K. A. Fischer, T. Sarmiento, K. G. Lagoudakis, Y. A. Kelaita, C. S. Mu\~{n}oz, E. del Valle, F. P. Laussy, and J. Vu\v{c}kovi\'{c}, Coherent generation of nonclassical light on chip via detuned photon blockade, Phys. Rev. Lett. \textbf{114}, 233601 (2015).
\bibitem{Peyronel572012}T. Peyronel, O. Firstenberg, Q.-Y. Liang, S. Hofferberth, A. V. Gorshkov, T. Pohl, M. D. Lukin, and V. Vuleti\'{c}, Quantum nonlinear optics with single photons enabled by strongly interacting atoms, Nature (London) \textbf{488}, 57 (2012).
\bibitem{Zhu0638422017}C. J. Zhu, Y. P. Yang, and G. S. Agarwal, Collective multiphoton blockade in cavity quantum electrodynamics, Phys. Rev. A \textbf{95}, 063842 (2017).
\bibitem{Lin0538502019}J. Z. Lin, K. Hou, C. J. Zhu, and Y. P. Yang, Manipulation and improvement of multiphoton blockade in a cavity-QED system with two cascade three-level atoms, Phys. Rev. A \textbf{99}, 053850 (2019).
\bibitem{Hou0638172019}K. Hou, C. J. Zhu, Y. P. Yang, and G. S. Agarwal, Interfering pathways for photon blockade in cavity QED with one and two qubits, Phys. Rev. A \textbf{100}, 063817 (2019).
\bibitem{Ebrahimi562023}M. S. Ebrahimi and M. B. Harouni, Parity-time symmetryenhanced simultaneous magnon and photon blockade in cavity magnonic system, J. Phys. B: At. Mol. Opt. Phys. \textbf{56}, 235501 (2023).
\bibitem{Lang2436012011}C. Lang, D. Bozyigit, C. Eichler, L. Steffen, J. M. Fink, A. A. Abdumalikov, M. Baur, S. Filipp, M. P. da Silva, A. Blais, and A. Wallraff, Observation of resonant photon blockade at microwave frequencies using correlation function measurements, Phys. Rev. Lett. \textbf{106}, 243601 (2011).
\bibitem{Hoffman0536022011}A. J. Hoffman, S. J. Srinivasan, S. Schmidt, L. Spietz, J. Aumentado, H. E. T¨¹reci, and A. A. Houck, Dispersive photon blockade in a superconducting circuit, Phys. Rev. Lett. \textbf{107}, 053602 (2011).
\bibitem{Faraon8592008}A. Faraon, I. Fushman, D. Englund, N. Stoltz, P. Petroff, and J. Vu\v{c}kovi\'{c}, Coherent generation of non-classical light on a chip via photon-induced tunnelling and blockade, Nat. Phys. \textbf{4}, 859 (2008).
\bibitem{Xu0438222014}X.W. Xu and Y. Li, Tunable photon statistics in weakly nonlinear photonic molecules, Phys. Rev. A \textbf{90}, 043822 (2014).
\bibitem{Gerace0318022014}D. Gerace and V. Savona, Unconventional photon blockade in doubly resonant microcavities with second-order nonlinearity, Phys. Rev. A \textbf{89}, 031803(R) (2014).
\bibitem{Tang92522015}J. Tang, W. Geng, and X. Xu, Quantum interference induced photon blockade in a coupled single quantum dot-cavity system, Sci. Rep. \textbf{5}, 9252 (2015).
\bibitem{Flayac0538102017}H. Flayac and V. Savona, Unconventional photon blockade, Phys. Rev. A \textbf{96}, 053810 (2017).
\bibitem{Wang2404022021}Y. Wang, W. Verstraelen, B. Zhang, T. C. H. Liew, and Y. D. Chong, Giant enhancement of unconventional photon blockade in a dimer chain, Phys. Rev. Lett. \textbf{127}, 240402 (2021).
\bibitem{Shen0437142024}H. Z. Shen, J. F. Yang, and X. X. Yi, Unconventional photon blockade with non-Markovian effects in driven dissipative coupled cavities, Phys. Rev. A \textbf{109}, 043714 (2024).
\bibitem{Zuo220202024}Zuo Y, Jiao Y-F, Xu X-W, Miranowicz A, Kuang L-M, Jing H, Chiral photon blockade in the spinning Kerr resonator, Opt. Express, \textbf{32} 22020 (2024).
\bibitem{Snijders0436012018}H. J. Snijders, J. A. Frey, J. Norman, H. Flayac, V. Savona, A. C. Gossard, J. E. Bowers, M. P. van Exter, D. Bouwmeester, and W. L\"{o}ffler, Observation of the unconventional photon blockade, Phys. Rev. Lett. \textbf{121}, 043601 (2018).
\bibitem{Vaneph0436022018}C. Vaneph, A. Morvan, G. Aiello, M. F¨¦chant, M. Aprili, J. Gabelli, and J. Est¨¨ve, Observation of the unconventional photon blockade in the microwave domain, Phys. Rev. Lett. \textbf{121}, 043602 (2018).
\bibitem{Hamsen1336042017}C. Hamsen, K. N. Tolazzi, T. Wilk, and G. Rempe, Two-Photon Blockade in an Atom-Driven Cavity QED System, Phys. Rev. Lett. \textbf{118}, 133604 (2017).
\bibitem{Wang20642020}D. Y. Wang, C. H. Bai, X. Han, S. Liu, S. Zhang, and H. F. Wang, Enhanced photon blockade in an optomechanical system with parametric amplification, Opt. Lett. \textbf{45}, 2604 (2020).
\bibitem{Shi82018}H.-Q. Shi, X.-T. Zhou, X.-W. Xu, and N.-H. Liu, Tunable phonon blockade in quadratically coupled optomechanical systems, Sci. Rep. \textbf{8}, 1 (2018).
\bibitem{Yuan222022}Z. Yuan, H. F. Wang and A. D. Zhu, Controllable photon blockade in double-cavity optomechanical system with Kerr-type nonlinearity, Quantum Inf. Process. \textbf{21}, 22 (2022).
\bibitem{Zheng0138042019}L.-L. Zheng, T.-S. Yin, Q. Bin, X.-Y. L\"{u}, and Y. Wu, Single-photon-induced phonon blockade in a hybrid spin-optomechanical system, Phys. Rev. A \textbf{99}, 013804 (2019).
\bibitem{Zhang117732021}W. Zhang, D. Y. Wang, C. H. Bai, T. Wang, S. Zhang, and H. F. Wang, Generation and transfer of squeezed states in a cavity magnomechanical system by two-tone microwave fields, Opt. Express \textbf{29}, 11773 (2021).
\bibitem{Huang1536012018}R. Huang, A. Miranowicz, J.-Q. Liao, F. Nori, and H. Jing, Nonreciprocal photon blockade, Phys. Rev. Lett. \textbf{121}, 153601 (2018).
\bibitem{Zhou0238382015}Y. H. Zhou, H. Z. Shen, and X. X. Yi, Unconventional photon blockade with second-order nonlinearity, Phys. Rev. A \textbf{92}, 023838 (2015).
\bibitem{Shen0638082015}H. Z. Shen, Y. H. Zhou, and X. X. Yi, Tunable photon blockade in coupled semiconductor cavities, Phys. Rev. A \textbf{91}, 063808 (2015).
\bibitem{Ferretti0250122012}S. Ferretti, V. Savona, and D. Gerace, Optimal antibunching in passive photonic devices based on coupled nonlinear resonators, New J. Phys. \textbf{15}, 025012 (2013).
\bibitem{Flayac0338362013}H. Flayac and V. Savona, Input-output theory of the unconventional photon blockade, Phys. Rev. A \textbf{88}, 033836 (2013).
\bibitem{Kyriienko0638052014}O. Kyriienko and T. C. H. Liew, Triggered single-photon emitters based on stimulated parametric scattering in weakly nonlinear systems, Phys. Rev. A \textbf{90}, 063805 (2014).
\bibitem{Liu23004222024}Q. H. Liu, G. C. Wang, T. Z. Luan, and H. Z. Shen, Atom Mediated Single-Photon Nonlinearity in a Quadratically Coupled Optomechanical System, Adv. Quantum Technol. \textbf{7}, 2300422 (2024).
\bibitem{Zhang23001872023}W. Zhang, S. Liu, S. Zhang, and H. F. Wang, Kerr-nonlinearity enhanced photon blockades via driving a Delta-type atom, Adv. Quantum Technol. \textbf{6}, 2300187 (2023).
\bibitem{Shen328352015}H. Z. Shen, Y. H. Zhou, H. D. Liu, G. C. Wang, and X. X. Yi, Exact optimal control of photon blockade with weakly nonlinear coupled cavities, Opt. Express \textbf{23}, 32835 (2015).
\bibitem{Lemonde0638242014}M. A. Lemonde, N. Didier, and A. A. Clerk, Antibunching and unconventional photon blockade with Gaussian squeezed states, Phys. Rev. A \textbf{90}, 063824 (2014).
\bibitem{Shen0238562018}H. Z. Shen, C. Shang, Y. H. Zhou, and X. X. Yi, Unconventional single-photon blockade in non-Markovian systems, Phys. Rev. A \textbf{98}, 023856 (2018).
\bibitem{Sarma0138262018}B. Sarma and A. K. Sarma, Unconventional photon blockade in three-mode optomechanics, Phys. Rev. A \textbf{98}, 013826 (2018).
\bibitem{Zhou0438192018}Y. H. Zhou, H. Z. Shen, X. Y. Zhang, and X. X. Yi, Zero eigenvalues of a photon blockade induced by a non-Hermitian hamiltonian with a gain cavity, Phys. Rev. A \textbf{97}, 043819 (2018).
\bibitem{Miranowicz0238092013}A. Miranowicz, M. Paprzycka, Y. X. Liu, J. Bajer, and F. Nori, Two-photon and three-photon blockades in driven nonlinear systems, Phys. Rev. A \textbf{87}, 023809 (2013).
\bibitem{Zhai276492019}C. L. Zhai, R. Huang, H. Jing, and L. M. Kuang, Mechanical switch of photon blockade and photon-induced tunneling, Opt. Express \textbf{27}, 27649 (2019).
\bibitem{Wang26042020}D. Y. Wang, C. H. Bai, X. Han, S. Liu, S. Zhang, and H. F. Wang, Enhanced photon blockade in an optomechanical system with parametric amplification, Opt. Lett. \textbf{45}, 2604 (2020).
\bibitem{Li0437022024}Y. C. Li, Z. H. Yao, and H. Yang, One-photon and two-photon blockades in a four-wave-mixing system embedded with an atom, Phys. Rev. A \textbf{109}, 043702 (2024).
\bibitem{Qiao0537022024}X. F. Qiao, Z. G. Yao, and H. Yang, Strongly enhanced photon-pair blockade with three-wave mixing by quantum interference, Phys. Rev. A \textbf{110}, 053702 (2024).
\bibitem{Feng0435092021}L.-J. Feng and S.-Q. Gong, Two-photon blockade generated and enhanced by mechanical squeezing, Phys. Rev. A \textbf{103}, 043509  (2021).
\bibitem{Bin0438582018}Q. Bin, X.-Y. L\"{u}, S.-W. Bin, and Y. Wu, Two-photon blockade in a cascaded cavity-quantum-electrodynamics system, Phys. Rev. A \textbf{98}, 043858 (2018).
\bibitem{Kudlaszyk0538572019}A. Kowalewska-Kud{\l}aszyk, S. I. Abo, G. Chimczak, J. Pe\v{r}ina, Jr., F. Nori, and A. Miranowicz, Two-photon blockade and photon-induced tunneling generated by squeezing, Phys. Rev. A \textbf{100}, 053857 (2019).
\bibitem{Kamal3112011}A. Kamal, J. Clarke, and M. H. Devoret, Noiseless nonreciprocity in a parametric active device, Nat. Phys. \textbf{7}, 311 (2011).
\bibitem{Shen17972018}Z. Shen, Y.-L. Zhang, Y. Chen, F.-W. Sun, X.-B. Zou, G.-C. Guo, C.-L. Zou, and C.-H. Dong, Reconfigurable optomechanical circulator and directional amplifier, Nat. Commun. \textbf{9}, 1797 (2018).
\bibitem{Malz0236012018}D. Malz, L. D. T¨®th, N. R. Bernier, A. K. Feofanov, T. J. Kippenberg, and A. Nunnenkamp, Quantum-limited directional amplifiers with optomechanics, Phys. Rev. Lett. \textbf{120}, 023601 (2018).
\bibitem{Sounas7742017}D. L. Sounas and A. Al\'{u}, Non-reciprocal photonics based on time modulation, Nat. Photon. 11, 774 (2017).
\bibitem{Jiao1436052020}Y.-F. Jiao, S.-D. Zhang, Y.-L. Zhang, A. Miranowicz, L.-M. Kuang, and H. Jing, Nonreciprocal optomechanical entanglement against backscattering losses, Phys. Rev. Lett. \textbf{125}, 143605 (2020).
\bibitem{Jiao0640082022}Y.-F. Jiao, J.-X. Liu, Y. Li, R. Yang, L.-M. Kuang, and H. Jing, Nonreciprocal enhancement of remote entanglement between nonidentical mechanical oscillators, Phys. Rev. Appl. \textbf{18}, 064008 (2022).
\bibitem{Ren11252022}Y.-L. Ren, Nonreciprocal optical¨Cmicrowave entanglement in aspinning magnetic resonator, Opt. Lett. \textbf{47}, 1125 (2022).
\bibitem{Chen0241052023}J. Chen, X.-G. Fan,W. Xiong, D.Wang, and L. Ye, Nonreciprocal entanglement in cavity-magnon optomechanics, Phys. Rev. B \textbf{108}, 024105 (2023).
\bibitem{Xu0535012021}Y. Xu, J. Y. Liu, W. Liu, and Y. F. Xiao, Nonreciprocal phonon laser in a spinning microwave magnomechanical system, Phys. Rev. A \textbf{103}, 053501 (2021).
\bibitem{Huang33112022}K. W. Huang, Y. Wu, and L. G. Si, Parametric-amplification-induced nonreciprocal magnon laser, Opt. Lett. \textbf{47}, 3311 (2022).
\bibitem{Huang1042012023}K. W. Huang, X. Wang, Q. Y. Qiu, L. Wu, and H. Xiong, Nonreciprocal phonon laser in an asymmetric cavity with an atomic ensemble, Chin. Phys. Lett. \textbf{40}, 104201 (2023).
\bibitem{Xu52762021}Y. J. Xu and J. Song, Nonreciprocal magnon laser, Opt. Lett. \textbf{46}, 5276 (2021).
\bibitem{He435062023}X. W. He, Z. Y. Wang, X. Han, S. Zhang, and H. F. Wang, Parametrically amplified nonreciprocal magnon laser in a hybrid cavity optomagnonical system, Opt. Express \textbf{31}, 43506 (2023).
\bibitem{Wang49872024}Z. Y. Wang, X. W. He, X. Han, H. F. Wang, and S. Zhang, Nonreciprocal PT-symmetric magnon laser in spinning cavity optomagnonics, Opt. Express \textbf{32}, 4987 (2024).
\bibitem{Jing14242018}H. Jing, H. L\"{u}, S. K. \"{O}zdemir, T. Carmon, and F. Nori, Nanoparticle sensing with a spinning resonator, Optica \textbf{5}, 1424 (2018).
\bibitem{Mirza255152019}I. M. Mirza, W. Ge, and H. Jing, Optical nonreciprocity and slow light in coupled spinning optomechanical resonators, Opt. Express \textbf{27}, 25515 (2019).
\bibitem{Peng0335072023}M. Peng, H. Zhang, Q. Zhang, T.-X. Lu, I. M. Mirza, and H. Jing, Nonreciprocal slow or fast light in anti-PT-symmetric optomechanics, Phys. Rev. A \textbf{107}, 033507 (2023).
\bibitem{Li0535222021}B. Li, \c{C}. K. \"{O}zdemir, X.-W. Xu, L. Zhang, L.-M. Kuang, and H. Jing, Nonreciprocal optical solitons in a spinning Kerr resonator, Phys. Rev. A \textbf{103}, 053522 (2021).
\bibitem{Jing0337072021}Y.-W. Jing, H.-Q. Shi, and X.-W. Xu, Nonreciprocal photon blockade and directional amplification in a spinning resonator coupled to a two-level atom, Phys. Rev. A \textbf{104}, 033707 (2021).
\bibitem{Zhang2403132023}W. Zhang, T. Wang, S. Liu, S. Zhang, and H.-F. Wang, Nonreciprocal photon blockade in a spinning resonator coupled to two two-level atoms, Sci. China Phys. Mech. Astron. \textbf{66}, 240313 (2023).
\bibitem{Yuan0535262024}N. Yuan, S.-Y. Li, N. Wang, T.-T. Dong, and A.-D. Zhu, Phase-modulated nonreciprocal photon blockade and transmission via optomechanically induced Kerr nonlinearity, Phys. Rev. A \textbf{109}, 053526 (2024).
\bibitem{Xue44242020}W. S. Xue, H. Z. Shen, and X. X. Yi, Nonreciprocal conventional photon blockade in driven dissipative atom-cavity, Opt. Lett. \textbf{45}, 4424 (2020).
\bibitem{Li6302019}B. Li, R. Huang, X. Xu, A. Miranowicz, and H. Jing, Nonreciprocal unconventional photon blockade in a spinning optomechanical system, Photon. Res. \textbf{7}, 630 (2019).
\bibitem{Hou51452025}R. Hou, W. Zhang, X. Han, H.-F. Wang and S. Zhang, Nonreciprocal unconventional photon blockade in a spinning microwave magnomechanical system, Sci. Rep. \textbf{15}, 5145 (2025).
\bibitem{Shen0138262020}H. Z. Shen, Q. Wang, J. Wang, and X. X. Yi, Nonreciprocal unconventional photon blockade in a driven dissipative cavity with parametric amplification, Phys. Rev. A \textbf{101}, 013826 (2020).
\bibitem{Xia0637132021} X. W. Xia, X. Q. Zhang, J. P. Xu, H. Z. Li, Z. Y. Fu, and Y. P. Yang, Giant nonreciprocal unconventional photon blockade with a single atom in an asymmetric cavity, Phys. Rev. A \textbf{104}, 063713 (2021).
\bibitem{Xia79072022}X. W. Xia, X. Q. Zhang, J. P. Xu, H. Z. Li, Z. Y. Fu, and Y. P. Yang, Improvement of nonreciprocal unconventional photon blockade by two asymmetrical arranged atoms embedded in a cavity, Opt. Express \textbf{30}, 7907 (2022).
\bibitem{Wang640032021}J. Wang, Q. Wang, and H. Z. Shen, Nonreciprocal unconventional photon blockade with spinning atom-cavity, Europhys. Lett. \textbf{134}, 64003 (2021);
\bibitem{Xu1432020}X.-W. Xu, Y. Zhao, H. Wang, H. Jing, and A. Chen, Quantum nonreciprocality in quadratic optomechanics, Photon. Res. \textbf{8}, 143 (2020).
\bibitem{Shang1152022021}X. Shang, H. Xie, and X. M. Lin, Nonreciprocal photon blockade in a spinning optomechanical resonator, Laser Phys. Lett. \textbf{18}, 115202 (2021)
\bibitem{Liu128472023}Y.-M. Liu, J. Cheng, H.-F. Wang, and X. Yi, Nonreciprocal photon blockade in a spinning optomechanical system with nonreciprocal coupling, Opt. Express \textbf{31}, 12847 (2023).
\bibitem{Gu0437222022}W.-J. Gu, L. Wang, Z. Yi, and L.-H. Sun, Generation of nonreciprocal single photons in the chiral waveguide-cavity-emitter system, Phys. Rev. A \textbf{106}, 043722 (2022).
\bibitem{Liu0637012023}Y. M. Liu, J. Cheng, H. F. Wang, and X. Yi, Simultaneous nonreciprocal conventional photon blockades of two independent optical modes by a two-level system, Phys. Rev. A \textbf{107}, 063701 (2023).
\bibitem{Gou0437232023}C. Gou and X. Hu, Simultaneous nonreciprocal photon blockade in two coupled spinning resonators via Sagnac-Fizeau shift and parametric amplification, Phys. Rev. A \textbf{108}, 043723 (2023).
\bibitem{Zhang0237232024}W. Zhang, R. Hou, T. Wang, S. Liu, S. Zhang, and H.-F. Wang, Simultaneous nonreciprocal photon blockade via directional parametric amplification, Phys. Rev. A \textbf{110}, 023723 (2024).
\bibitem{Malykin12292000}G. B. Malykin, The Sagnac effect: correct and incorrect explanations, Phys.-Usp. \textbf{43}, 1229 (2000).
\bibitem{Scully1997}M. O. Scully and M. S. Zubairy, \textit{Quantum Optics} (Cambridge University Press, Cambridge, 1997).
\bibitem{Walls1994}D. F. Walls and G. J. Milburn, \textit{Quantum Optics} (Springer- Verlag, Berlin, 1994).
\bibitem{Glauber25291963}R. J. Glauber, The quantum theory of optical coherence, Phys. Rev. \textbf{130}, 2529 (1963).
\bibitem{Vahala8392003}K. J. Vahala, Optical microcavities, Nature (London) \textbf{424}, 839 (2003).
\bibitem{Spillane0138172005}S. M. Spillane, T. J. Kippenberg, K. J. Vahala, K. W. Goh, E. Wilcut, and H. J. Kimble, Ultrahigh-Q toroidal microresonators for cavity quantum electrodynamics, Phys. Rev. A \textbf{71}, 013817 (2005).
\bibitem{Aspelmeyer13912014}M. Aspelmeyer, T. J. Kippenberg, and F. Marquardt, Cavity optomechanics, Rev. Mod. Phys. \textbf{86}, 1391 (2014).
\bibitem{Peng3942014}B. Peng, S. K. \"{O}zdemir, F. Lei, F. Monifi, M. Gianfreda, G. L. Long, S. Fan, F. Nori, C. M. Bender, and L. Yang, Parity-time symmetric whispering-gallery microcavities, Nat. Phys. \textbf{10}, 394 (2014).


\end{references}
\end{document}